\documentclass[
12pt,
3p,
review,
]{elsarticle}
\usepackage{lineno,hyperref}
\modulolinenumbers[1]
\usepackage{graphicx,graphics}
\usepackage{placeins}
\usepackage{amsmath}
\usepackage{amssymb}
\usepackage{mathtools}
\usepackage{caption}
\usepackage{array}
\usepackage{subcaption}
\usepackage{lineno,hyperref}
\usepackage{color}
\usepackage{soul}
\usepackage{dcolumn}
\usepackage{bm}
\usepackage{todonotes}
\usepackage[utf8]{inputenc}
\usepackage[english]{babel}
\usepackage{libertine}
\usepackage{upgreek}
\usepackage{gensymb}

\usepackage{layouts}

\modulolinenumbers[5]

\newcommand{\vphi}{\mbox{{\bm{$\phi$}}}}

\newcommand{\Ms}{M_\text{s}}

\newcommand{\vv}[1]{\boldsymbol{#1}}
\newcommand{\n}{\vv{\nabla}}
\renewcommand{\l}{\mathopen{}\mathclose\bgroup\left}
\renewcommand{\r}{\aftergroup\egroup\right}
\newcommand{\veps}{\mbox{{\bm{$\epsilon$}}}}
\newcommand{\diff} [1]{\mathrm{d}{#1}} 
\newcommand{\phia}{\phi_{\alpha}}
\newcommand{\phib}{\phi_{\beta}}
\newcommand{\gab}{\gamma_{\alpha \beta}}
\newcommand{\gabd}{\gamma_{\alpha \beta \delta}}
\newcommand{\qab}{q_{\alpha \beta}}
\newcommand{\C}{\vv{\mathcal{C}}}

\let\originaleps=\epsilon
\let\epsilon=\varepsilon
\let\varepsilon=\originaleps
\usepackage{stmaryrd}
\newcommand{\ljump}{\llbracket}
\newcommand{\rjump}{\rrbracket}
\renewcommand{\S}{\vv{\mathcal{S}}}

\newcommand{\pace}{{\textsc{Pace3D}}}

\definecolor{mydark_blue}{RGB}{0, 0, 139}
\definecolor{myblue}{RGB}{0, 0, 255}
\definecolor{mycyan}{RGB}{0, 255, 255}  
\definecolor{mygreen}{RGB}{0, 255, 0}
\definecolor{myyellow}{RGB}{255, 255, 0}
\definecolor{myred}{RGB}{255, 0, 0}
\definecolor{mydark_red}{RGB}{139, 0, 0}
\definecolor{myblack}{RGB}{0, 0, 0}

\definecolor{BRY_1}{RGB}{  0,  0,255}
\definecolor{BRY_2}{RGB}{127,  0,127}
\definecolor{BRY_3}{RGB}{255,  0,  0}
\definecolor{BRY_4}{RGB}{255,127,  0}
\definecolor{BRY_5}{RGB}{255,255, 85}

\newcommand{\T}{\mathsf{\!T}}
\newcommand{\TT}{\vv{\mathcal{T}}}

\newcommand{\laplace}{\Delta}
    
\journal{Computational Material Science}

\begin{document}

\begin{frontmatter}
\title{Phase-field analysis of quenching and partitioning in a polycrystalline Fe-C system under constrained-carbon equilibrium condition}

\author[mymainaddress]{P G Kubendran Amos\corref{mycorrespondingauthor}\fnref{fn1}}
\cortext[mycorrespondingauthor]{Prince Gideon Kubendran Amos}
\ead{prince.amos@kit.edu}

\author[mymainaddress,mysecondaryaddress]{Ephraim Schoof\fnref{fn1}}
\author[mysecondaryaddress]{Nick Streichan}
\author[mymainaddress,mysecondaryaddress]{Daniel Schneider}
\author[mymainaddress,mysecondaryaddress]{Britta Nestler}

\fntext[fn1]{The authors contributed equally.}

\address[mymainaddress]{Institute of Applied Materials (IAM-CMS), Karlsruhe Institute of Technology (KIT),\\
Strasse am Forum 7, 76131 Karlsruhe, Germany
}
\address[mysecondaryaddress]{Institute of Digital Materials Science (IDM), Karlsruhe University of Applied Sciences,\\
Moltkestr. 30, 76133 Karlsruhe, Germany}

\begin{abstract} 

Mechanical properties of steels are significantly enhanced by retained austenite.
Particularly, it has been shown that a recently developed heat-treatment technique called Quenching and Partitioning (Q\&P) stabilises austenite effectively.
In the present work, the phase-field approach is adopted to simulate the phase transformation and carbon diffusion, which respectively accompanies the quenching and partitioning process of the polycrystalline Fe-C system. 
By incorporating the chemical driving-force from the CALPHAD database, the elastic phase-field model, which recovers the sharp-interface solutions, simulates the martensite ($\alpha'$) transformation at three different quenching temperatures. 
The resulting martensite volume-fractions are in complete agreement with the analytical predictions.
For the first time, in this study, the constrained carbon equilibrium (CCE) condition is introduced in the polycrystalline set-up to yield the predicted partitioning endpoints. 
Under the CCE condition, the carbon partitioning in two alloys of varying composition is analysed through the phase-field model which employs chemical potential as the dynamic variable. 
The volume fraction and distribution of retained austenite is determined from the carbon distribution and its temporal evolution during the partitioning is investigated. 
It is identified that in the initial stages of partitioning carbon gets accumulated in the austenite ($\gamma$) along the $\gamma\alpha'$-interface, owing to the substantial difference in the diffusivities and CCE endpoints.
This accumulation stabilises the austenite adjacent to the interface.
However, depending on the martensite volume-fraction and the alloy composition, the evolution of the stabilised austenite varies.
Furthermore, the influence of the phase distribution on the kinetics of the temporal evolution of retained austenite is elucidated. 

\end{abstract}

\begin{keyword}
\texttt Quenching and partitioning, Constrained carbon equilibrium, Martensite transformation, Carbon partitioning, Retained austenite, Multiphase-field simulation
\end{keyword}

\end{frontmatter}


\section{Introduction}

Material research primarily directed towards optimising the energy consumption are increasingly favoured, due to its overt implications. 
Particularly in the automotive industries, attempts are extensively made to introduce fuel-efficient materials with enhanced mechanical properties. 
In order to achieve the desired properties, the microstructures of the materials are appropriately transformed through meticulously devised heat treatment techniques.
Owing to the toughness and strength, which is respectively rendered by austenite and martensite, the combination of these two phases is preferred in steels~\cite{bhadeshia1979tempered,ojima1998application}.
Although, the microstructure consisting of martensite and austenite can be attained by conventional quenching, the low stability of the retained austenite introduces a change in the predicted volume fraction of the constituent phases, thereby digressing from the expected properties~\cite{thomas1978retained}. 
Alternatively, a heat treatment technique, referred to as quenching and partitioning (Q\&P), is thus 	employed as a unique approach to yield a microstructure of stable austenite in a matrix of martensite~\cite{speer2003carbon}. 
From the outset, this approach has proven to be an efficient technique in enhancing the properties of steels used in automotive applications~\cite{speer2004partitioning,matlock2003application}.

\begin{figure}
    \centering
      \begin{tabular}{@{}c@{}}
      \includegraphics[width=0.85\textwidth]{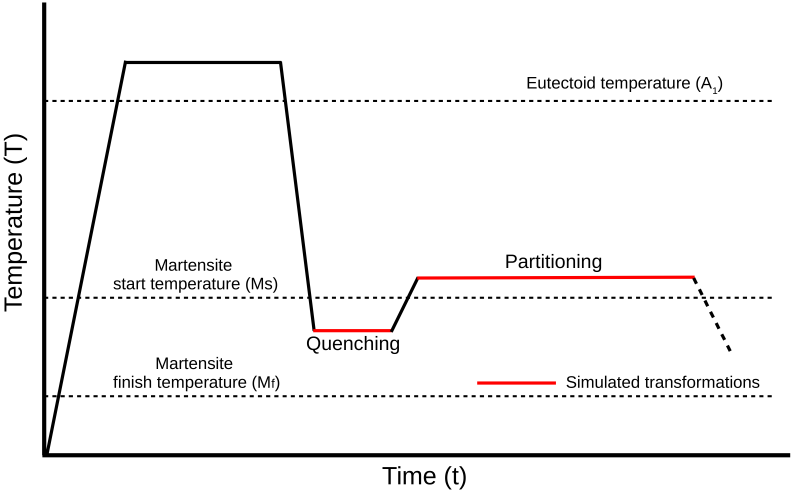}
    \end{tabular}
    \caption{ A schematic illustration of the thermal cycle adopted for the quenching and partitioning (Q\&P) treatment. The regime of this cycle simulated in the present work is highlighted.
    \label{fig:thermal_cycle}}
\end{figure}

A schematic representation of the heat treatment cycle involved in the Q\&P technique is presented in Fig~\ref{fig:thermal_cycle}. 
The processing begins with the austenization of the steel, although in some instances partial-austenization or intercritical annealing is also adopted~\cite{santofimia2009microstructural}. 
This austenized steel is then quenched to a temperature between martensite start and finish temperature, $\Ms$ and $M_\text{f}$ temperature, respectively. 
The quenching temperature is chosen to achieve the required volume fraction of martensite.
During quenching, the pre-determined amount of austenite transforms to primary martensite.
The quenching is subsequently followed by the partitioning.
In this stage, the steel is raised to relatively higher temperature, within the austenizing temperature. 
Owing to the increased diffusivity, the carbon from the supersaturated martensite diffuses into the austenite. 
The increase in the carbon concentration lowers the $\Ms$ temperature of the austenite, and thereby enhances its stability.
Upon partitioning, the steel is quenched to room temperature and the austenite which are not sufficiently enriched with carbon transform to secondary martensite.

When compared with the tempering of martensite, the partitioning in the Q\&P intends to preclude the carbide formation and the decomposition of austenite into ferrite or bainite~\cite{santofimia2011overview}. 
The absence of carbide facilitates the accumulation of carbon in the austenite which in turn improves its stability. 
Furthermore, by considering the interface between austenite and martensite to be stationary~\cite{toji2014atomic}, it is postulated that a specific equilibrium condition, called as constrained para- or carbon equilibrium (CCE), is established between the phases~\cite{speer2003carbon,hillert2004definitions}.
In other words, since the volume fraction of the phases during partitioning, though consistent to the quenching temperature, deviates significantly from the tie-line based estimation of the phase diagram, a considerable deviation from the equilibrium composition is predicted at the endpoints of the partitioning. 
Additionally, by assuming the lack of migration of any substitutional atoms, the composition of the phases at the end of the partition is calculated based on the activity of carbon.

Early investigations immediately following the introduction of the Q\&P are collectively presented in Refs.~\cite{clarke2008carbon,speer2015critical}.
In these studies, the partitioning of carbon from the supersaturated martensite to austenite is asserted by the increase in the volume fraction of the retained austenite.
Recently, however, the partitioning of carbon has been substantiated by in-situ neutron diffraction~\cite{bigg2013real,bigg2011dynamics}, transmission electron microscopy and atomic probe tomography~\cite{toji2014atomic,toji2015carbon}. 
In addition to the experimental studies, theoretical treatments have been involved to understand the intricacies of the carbon partitioning and to optimise the heat treatment cycle.
By employing the Koistinen-Marburger relation~\cite{koistinen1959general}, an optimum quenching temperature, which precludes the formation of secondary martensite, is calculated~\cite{speer2004partitioning}.
This approach is further extended by introducing the kinetics of carbon partitioning and the final volume-fraction of the austenite is ascertained~\cite{clarke2009influence,santofimia2009influence}.
Furthermore, motivated by the experimental observation~\cite{zhong2006interface}, the behaviour of the interface during the partitioning of carbon is also analytically investigated~\cite{speer2007influence,santofimia2008model}.
Despite the critical claims, owing to the one-dimensional nature of the set-up and the idealised partitioning conditions, inadequacies of these analyses are often conceded.

With the availability of the computational resources, the numerical simulations have been increasingly involved in enhancing the understanding of the complex microstructural transformations.
Phase-field modelling is one such computational approach which is particularly gaining ground in simulating thermodynamically-consistent evolution of large-scale polycrystalline systems~\cite{chen2002phase,steinbach2013phase}.
When compared to sharp interface models, the phase-field approach obviates the need for the strenuous tracking of the interface in a polycrystalline system, by assigning order parameters to the individual phases and grains~\cite{perumal2017phase,amos2018phase2,mittnacht2018understanding}.
The smooth transition between the order parameters of the bulk phases replaces the sharp interface with a well-defined diffuse region, which is subsequently treated as the interface. 
In spite of the introduction of the diffuse interface,  through the asymptotic analysis it is proved that the sharp-interface solutions, including the physical laws, are recovered at the interfaces and multiple junctions~\cite{caginalp1989stefan,garcke1998anisotropic}. 
Furthermore, the advancements in the phase-field models have enabled the incorporation of appropriate thermodynamic data, for example from the CALPHAD database, thereby enhancing the quantitative nature of the simulations~\cite{steinbach2007calphad}.
Phase-field modelling has already been adopted to analyse the carbon partitioning during the Q\&P technique in a polycrystalline system~\cite{takahama2012phase}.
Focusing on the low-carbon transformation-induced plastic (TRIP) steel, and extending the analytical treatment of the interface migration~\cite{santofimia2008model}, the theoretical study employs the phase-field approach to understand the diffusion pathways of carbon during partitioning and to investigate the behaviour of the interface.
Accordingly, it is identified that the carbon diffuses through the ferrite, in addition to the direct partitioning from supersaturated martensite to austenite, in an intercritically annealed steel. 
Furthermore, it is shown that the interface is not stationary but migrates governed by the difference in the free energy. 
The subsequent phase-field study analyses the Q\&P in a completely austenized steel, both in 2- and 3-dimensions, while considering acicular ferrite as martensite~\cite{mecozzi2016phase}.
Both existing phase-field studies consider the equi-partitioning of carbon which deviates significantly from the proposed constrained carbon equilibrium (CCE) condition.
Since the equilibrium composition of the phases differs substantially from the partitioning endpoints under CCE, the resulting volume-fraction of the retained austenite and its corresponding kinetics is expected to vary.

In the present work, thermodynamically-consistent phase-field models are employed to simulate the quenching and partitioning of fully austenized polycrystalline binary Fe-C steel.
Motivated by the experimental observations~\cite{toji2014atomic,toji2015carbon}, which indicate that the interface is largely stationary during the partitioning, any phase transformation accompanying the diffusion of carbon from martensite to austenite is considered marginal and thus averted.
Moreover, the constrained carbon equilibrium (CCE) condition is incorporated in this theoretical study, and the kinetics of the carbon partitioning is captured by introducing thermodynamically appropriate parameters.

\section{Model and simulation set-up}

\subsection{Multiphase-field model for quenching}\label{sec:quench}

In the Q\&P processing, quenching displacively decomposes austenite to martensite.
This diffusionless transformation in a polycrystalline system is simulated by employing an elastic multiphase-field model.
Although a comprehensive description of the model is presented elsewhere, Refs.~\cite{schneider2017small,schoof2017multiphase}, a concise elucidation is rendered in this section. 

In a polycrystalline system, the order parameter or phase field, which is introduced in addition to the other thermodynamic variables like concentration and temperature, is treated as an $N$-tuple variable, $\vv{\phi}(\vv{x}, t)={(\phi_{1}(\vv{x}, t),\phi_{2}(\vv{x}, t)....,\phi_{\alpha}(\vv{x}, t),\phi_{\beta}(\vv{x}, t),...,\phi_{N})}$, where $N$ corresponds to the number of phases or grains in a polycrystalline system, respectively.
Furthermore, $\phi_{\alpha}(\vv{x}, t)$ is the state variable which assumes $\phi_{\alpha}(\vv{x}, t)=1$ in the bulk region $\alpha$ and turns $\phi_{\alpha}(\vv{x}, t)=0$ outside.
Correspondingly, the phase-field variables represent the volume fraction of the individual phases or grains.

In the phase-field modelling, the transformation is dictated by the phenomenological change in the thermodynamically devised functional. 
In the present model, a Ginzburg-Landau type free-energy functional $\mathcal{F}(\vphi, \n \vphi, \bar{\veps})$, which consists of interfacial, elastic and chemical contributions, is considered~\cite{ginzburg1950theory}.
This free-energy functional is expressed as
\begin{align}\label{eq:functional}
  \mathcal{F}(\vphi, \n \vphi, \bar{\veps}) &= \mathcal{F}_\text{intf}(\vphi, \n \vphi) + \mathcal{F}_\text{el}(\vphi, \bar{\veps}) + \mathcal{F}_\text{chem}(\vphi) \\
  &= \int_V \bar{W}_\text{intf}(\vphi, \n \vphi) + \bar{W}_\text{el}(\vphi, \bar{\veps}) + \bar{W}_\text{chem}(\vphi) \diff{V},\nonumber
\end{align}
where the interfacial free-energy density is the summation of the gradient and the potential energy density,
$\bar{W}_\text{intf}(\vphi, \n \vphi) = \varepsilon a(\vv \phi, \n \vv{\phi}) + \omega(\vv{\phi})/\varepsilon$.
The width of the diffuse interface is dictated by the length scale parameter $\epsilon$.
The gradient energy density involved in the interface contribution is expressed as
\begin{align}\label{eq:a}
  \epsilon a(\vphi, \n \vphi) = \epsilon \sum_{\alpha, \beta > \alpha} \gab {|\qab|}^2,
\end{align}
where $a(\vphi, \n \vphi)$ imparts a form to the interface energy density $\gamma_{\alpha\beta}$, and the normal vector to the $\alpha\beta$-interface is $\qab = {\phi_{\alpha} \bm{\nabla}{\phi}_{\beta}} - {\phi_{\beta} \bm{\nabla}{\phi}_{\alpha}}$.
The potential energy density, formulated as an obstacle-type potential, reads
\begin{equation}\label{eq:potential_energy}
 \frac{1}{\epsilon}\omega(\vphi)= 
 \begin{cases}
 \frac{16}{\epsilon\pi^2} \underset{\alpha,\beta>\alpha}{\sum} \gab \phia \phib + \frac{1}{\epsilon}\underset{\alpha,\beta >\alpha,\delta >\beta}{\sum} \gamma_{\alpha \beta \delta} \phia \phib \phi_{\delta},& \text{if} ~\vphi \in \mathcal{G}\\
 \infty & \text{otherwise},
 \end{cases}
\end{equation}
where the Gibbs simplex $\mathcal{G}=\left\{ \vv{\phi} \in \mathbb{R}^N: \sum_\alpha \phi_\alpha = 1, \phi_\alpha \geq 0 \right\}$ assigns sharp minima to the bulk regions of phases, or grains respectively.
The higher-order term $\gamma_{\alpha \beta \delta} \phia \phib \phi_{\delta}$, in Eqn.~\ref{eq:potential_energy}, prevents the formation of non-physical spurious phase-field contributions in binary interface regions.
The analytical consistency and computational efficiency of adopting the obstacle-type potential is elucidated in Refs.~\cite{garcke1999multiphase,oono1988study,hotzer2016calibration}.

The elastic and chemical contribution from the bulk phases are expressed as the interpolation of phase-dependent free-energy densities,
\begin{align}
 \bar{W}_{\text{el}}(\vphi, \bar{\veps}) &=\sum_\alpha \phia W^\alpha_{\text{el}}(\vv\epsilon^\alpha)
\end{align}
and
\begin{align}\label{eq:bar_W_el}
\bar{W}_\text{chem}(\vphi) &= \sum_\alpha \phia W^\alpha_\text{chem}.
\end{align}
From the variation of $\mathcal{F}(\vphi, \n \vphi, \bar{\veps})$, the phase-field evolution is written as
\begin{align}\label{eq:phase_field_evolution}
  \frac{\partial \phia}{\partial t} &= - \frac{1}{\tilde{N}\varepsilon} \sum^{\tilde{N}}_{\beta \neq \alpha} M_{\alpha \beta} \bigg[\frac{\delta \mathcal{F}_\text{intf}}{\delta \phia} + \varepsilon\hat{a}(\phia, \nabla\phia) - \frac{\delta \mathcal{F}_\text{intf}}{\delta \phib} - \varepsilon\hat{a}(\phib, \nabla\phib) \\
  &- \frac{8 \sqrt{\phia \phib}}{\pi} \l( \Delta^{\alpha \beta}_\text{chem} + \Delta^{\alpha \beta}_\text{el} \r) \bigg] + \zeta \nonumber.
\end{align}
In the above Eqn.~\ref{eq:phase_field_evolution}, $\tilde{N}$ $(\leq N)$ is the number of active phase-fields and $M_{\alpha \beta}$ denotes the mobility of the $\alpha\beta$-interface.
Furthermore, the nucleation is modelled through an additional noise term~$\zeta$, which is active only in interfacial regions.
Since the nucleation of the martensite is not a critical aspect of the present analysis, it is not extensively discussed here.
However, the readers are directed to Ref.~\cite{schoof2017multiphase} for a comprehensive understanding on the formulation and consistency of the present nucleation approach.

Apart from the nucleation, the capillarity effect plays a subordinate role in martensite transformation when compared to the chemical and elastic driving forces.
Therefore, the stability of the interface is retained through the isotropic gradient energy density term
$
\varepsilon\hat{a}(\phia, \nabla\phia) = - \varepsilon\gamma^c_\alpha \left(\laplace \phia -|\nabla\phia| \nabla \cdot \left(\nabla \phia/|\nabla \phia|\right)\right),
$
which is scaled using an interface parameter, $\gamma^c_\alpha$~\cite{schoof2017multiphase}.
In Eqn.~\ref{eq:phase_field_evolution}, the variational derivatives of the chemical and elastic driving forces, $\l(\delta / \delta \phib - \delta / \delta \phia \r) \mathcal{F}_\text{chem}$ and  $\l(\delta / \delta \phib - \delta / \delta \phia \r) \mathcal{F}_\text{el}$, are represented as $\Delta^{\alpha \beta}_\text{chem}$ and $\Delta^{\alpha \beta}_\text{el}$, respectively.
The stresses and driving forces, contributing to the elastic free-energy density, are solved by incorporating the mechanical jump conditions~\cite{schneider2015phase, schneider2017small}.

Using the homogenised normal vector $\vv{n}$, which is determined from the scalar field through
\begin{align}\label{eq:n_aus_M}
M(\vv\phi)=\sum_{\alpha<\beta}\phia\phib \quad \Rightarrow \quad \vv n(M(\vv\phi)) = \frac{\nabla M(\vv\phi)}{|\nabla M(\vv\phi)|},
\end{align}
the stresses and strains are expressed in an orthonormal basis $\vv{B}=\{\vv{n},\vv{t},\vv{s}\}$.
In the Voigt notation, these stresses and strains are written as 
\begin{align}\label{eq:define_sigma_epsilon_n_t}
  \begin{split}
  \vv\sigma_B^\alpha(\vv{n}) :=&{\Big(
  \sigma_{nn},\sigma_{nt},\sigma_{ns},
  \sigma_{tt}^\alpha,\sigma_{ss}^\alpha,\sigma_{ts}^\alpha \Big)}^\T={\l(\vv\sigma_n, \vv\sigma_t^\alpha \r)}^\T \\
   \vv\epsilon_B^\alpha(\vv{n}) :=&{\Big(
  \epsilon_{nn}^\alpha,2 \epsilon_{nt}^\alpha,2 \epsilon_{ns}^\alpha,
  \epsilon_{tt},\epsilon_{ss},2 \epsilon_{ts}\Big)}^\T=
  {\left(\vv\epsilon_n^\alpha,\vv\epsilon_t\right)}^\T.
  \end{split}
\end{align}
With $\ljump \cdot \rjump^{\alpha\beta} = (\cdot)^\beta -(\cdot)^\alpha $ representing the jump of the variable across the interface, according to the force balance and the Hadamard kinematic compatibility condition, the jump of $\vv{\sigma}_n$ and $\vv{\epsilon}_t $ respectively vanishes ($\ljump \vv\sigma_n \rjump = \vv 0$ and $\ljump \vv\epsilon_t \rjump = \vv 0$).
Therefore, the continuous contribution of the stresses and strains, for an infinitesimal deformation on a singular plane, can be
simplified as $\vv{\sigma}_n := {(\sigma_{nn},\sigma_{nt},\sigma_{ns})}$ and $\vv{\epsilon}_t := {(\epsilon_{tt},\epsilon_{ss},2 \epsilon_{ts})}$.
The discontinuous contributions of the stresses and strains are respectively summarized as $\vv{\sigma}^\alpha_t := {(\sigma_{tt}^\alpha,\sigma_{ss}^\alpha,\sigma_{ts}^\alpha)}$ and $\vv\epsilon_n^\alpha := {(\epsilon_{nn}^\alpha,2 \epsilon_{nt}^\alpha,2 \epsilon_{ns}^\alpha)}$.

The displacement-field gradient, using the Einstein summation convention, is written as $(\nabla \vv u )_{ij} = \partial u_i/\partial x_j$. 
Accordingly, the local strain reads $\vv\epsilon = (\vv\nabla\vv u + (\vv\nabla\vv u)^{T})/2$.

The stiffness tensor is formulated in the orthonormal basis $\vv{B}$.
For the ease of numerical treatment, this tensor is divided into blocks
\begin{align}\label{eq:C_nt}
\C^\text{v}_B &=
\l(\begin{array}{ccc|ccc}
\mathcal{C}_{nnnn} & \mathcal{C}_{nnnt} & \mathcal{C}_{nnns} & \mathcal{C}_{nntt} & \mathcal{C}_{nnss} & \mathcal{C}_{nnts}\\
\mathcal{C}_{ntnn} & \mathcal{C}_{ntnt} & \mathcal{C}_{ntns} & \mathcal{C}_{nttt} & \mathcal{C}_{ntss} & \mathcal{C}_{ntts}\\
\mathcal{C}_{nsnn} & \mathcal{C}_{nsnt} & \mathcal{C}_{nsns} & \mathcal{C}_{nstt} & \mathcal{C}_{nsss} & \mathcal{C}_{nsts}\\ \hline
\mathcal{C}_{ttnn} & \mathcal{C}_{ttnt} & \mathcal{C}_{ttns} & \mathcal{C}_{tttt} & \mathcal{C}_{ttss} & \mathcal{C}_{ttts}\\
\mathcal{C}_{ssnn} & \mathcal{C}_{ssnt} & \mathcal{C}_{ssns} & \mathcal{C}_{sstt} & \mathcal{C}_{ssss} & \mathcal{C}_{ssts}\\
\mathcal{C}_{tsnn} & \mathcal{C}_{tsnt} & \mathcal{C}_{tsns} & \mathcal{C}_{tstt} & \mathcal{C}_{tsss} & \mathcal{C}_{tsts}
\end{array} \r)
=:
\begin{pmatrix}
\C_{nn} & \C_{nt}\\
\C_{tn} & \C_{tt}
\end{pmatrix},
\end{align}
where  $\C_{nn}$ and $\C_{tt}$ are the symmetrical matrices of dimension $3 \times 3$.
Furthermore, the $3 \times 3$ matrices $\C_{nt}$ and $\C_{tn}$ satisfy the condition $\C_{tn} = \C^\T_{nt}$.
The compliance tensor $\vv{\mathcal{S}}^\alpha$ is constructed in a similar manner.

By exclusively considering continuous variables, the stresses are calculated by
\begin{align}\label{eq:sigma_B_multi}
\vv{\bar{\sigma}}_{B} = 
 \underbrace{
 \begin{pmatrix}
  - \bar{\TT}_{nn}^{-1}                & -\bar{\TT}_{nn}^{-1} \bar{\TT}_{nt} \\
  - \bar{\TT}_{tn} \bar{\TT}_{nn}^{-1} & \bar{\TT}_{tt} - \bar{\TT}_{tn} \bar{\TT}_{nn}^{-1} \bar{\TT}_{nt}
  \end{pmatrix} 
}_{\bar{\C}^\text{v}_B (\vv{\phi})}
 \begin{pmatrix} 
 \vv{\epsilon}_n \\ \vv{\epsilon}_t 
 \end{pmatrix}
  + \underbrace{
  \begin{pmatrix}
  \bar{\TT}_{nn}^{-1}                & \vv{O} \\
  \bar{\TT}_{tn} \bar{\TT}_{nn}^{-1} & - \vv{I}
  \end{pmatrix} 
 \begin{pmatrix} 
 \tilde{\vv{\chi}}_n \\ \tilde{\vv{\chi}}_t 
 \end{pmatrix}
 }_{\tilde{\vv{\sigma}}_{B}}.
\end{align}
In Eqn.~\ref{eq:sigma_B_multi}, the normal and tangential components of the inelastic strains $\tilde{\vv\epsilon}^\alpha$ are expressed as 
\begin{align}\label{eq:chi_n_t_poly}
  \tilde{\vv\chi}_n = \sum_\alpha \l(\tilde{\vv{\epsilon}}^\alpha_n + \TT^\alpha_{nt}\tilde{\vv{\epsilon}}^\alpha_t \r) \phia,  \quad   \tilde{\vv\chi}_t = \sum_\alpha \TT^\alpha_{tt} \tilde{\vv{\epsilon}}^\alpha_t \phia,
\end{align}
respectively. 
The proportionality matrix, which relates the continuous and discontinuous variables, is locally averaged and is written as
\begin{align}
\bar{\TT}_{nn} &:= \sum_\alpha \TT^\alpha_{nn} \phia := - \sum_\alpha {(\C^\alpha_{nn})}^{-1} \phia\label{eq:T_nn} \\ 
\bar{\TT}_{nt} &:= \sum_\alpha \TT^\alpha_{nt} \phia :=  \sum_\alpha {(\C^\alpha_{nn})}^{-1} \C^\alpha_{nt} \phia \label{eq:T_nt} \\
\bar{\TT}_{tt} &:= \sum_\alpha \TT^\alpha_{tt} \phia := 
\sum_\alpha \l(\C^\alpha_{tt} - \C^\alpha_{tn} {(\C^\alpha_{nn})}^{-1} \C^\alpha_{nt} \r) \phia.
\end{align}
The resulting stresses in the Voigt notations are transformed to the Cartesian coordinate system by $ \vv{\bar{\sigma}}^\text{v}(\vv{\phi})  = \bar{\C}^\text{v}(\vv{\phi}) \bar{\vv{\epsilon}}^\text{v} + \tilde{\vv\sigma}^\text{v}(\vv{\phi})$, where $\bar{\C}^\text{v}(\vv{\phi})$ and $ \tilde{\vv\sigma}^\text{v}(\vv{\phi})$ are respectively expressed as $\bar{\C}^\text{v}(\vv{\phi}) = \vv{M}_\epsilon^\T \C^\text{v}_B (\vv{\phi}) \vv{M}_\epsilon$ and $\tilde{\vv{\sigma}}^\text{v}(\vv{\phi}) = \vv{M}_\sigma^\T \tilde{\vv\sigma}_{B}(\vv{\phi})$. 
The transformation matrices $\vv{M}_\epsilon$ and $\vv{M}_\sigma$ are adopted from Ref.~\cite{schneider2015phase}. 
Subsequently, the displacement field $\vv{u}$ is ascertained by solving the momentum balance $\n \cdot \bar{\vv{\sigma}}(\vv{\phi}) = \vv{0}$.

The stresses in the bulk phase ($\alpha$) are calculated by $ \sigma_{ij} = (\C^\alpha[\vv\epsilon -\tilde{\vv\epsilon}^\alpha])_{ij} =\mathcal{C}_{ijkl}^\alpha(\epsilon_{kl} -\tilde{\epsilon}^\alpha_{kl}).$
The derivative of the elastic free-energy density reads
$\partial\bar{W}_\text{el}(\vv \phi, \vv{\epsilon}_B)/\partial\phia = \partial\sum_\alpha p^\alpha(\vv{\sigma}_n, \vv{\epsilon}_t) \phia/\partial\phia$, 
wherein $p^\alpha(\vv{\sigma}_n, \vv{\epsilon}_t)$ is expressed as
\begin{align}\label{eq:p_alpha}
p^\alpha(\vv{\sigma}_n, \vv{\epsilon}_t) 
&=\frac{1}{2} \left( 
\begin{pmatrix} 
\vv{\sigma}_n \\ \vv{\epsilon}_t 
\end{pmatrix} 
\cdot 
\begin{pmatrix} 
  \TT^\alpha_{nn} & \TT^\alpha_{nt} \\
  \TT^\alpha_{tn} & \TT^\alpha_{tt}   
\end{pmatrix}
\begin{pmatrix}
  \vv{\sigma}_n \\ \vv{\epsilon}_t
\end{pmatrix} 
  \right) -\left( 
\begin{pmatrix} 
  \vv{\sigma}_n \\ \vv{\epsilon}_t 
\end{pmatrix} 
  \cdot 
\begin{pmatrix} 
  \vv{I} & \TT^\alpha_{nt}        \\
  \vv{O} & \TT^\alpha_{tt}
\end{pmatrix}
\begin{pmatrix} 
  \tilde{\vv\epsilon}^\alpha_n \\ \tilde{\vv\epsilon}^\alpha_t 
\end{pmatrix} 
  \right) \\ 
  &+\frac{1}{2}\left( \tilde{\vv\epsilon}^\alpha_t \cdot \TT^\alpha_{tt} \tilde{\vv\epsilon}^\alpha_t \right).\nonumber
\end{align}
The proportionality matrix $\TT^\alpha$ in Eqn.~\ref{eq:p_alpha} consists of the components
\begin{align}\label{eq:T_parts_multi}
\TT^\alpha_{nn}  &:= -  \S^\alpha_{nn} \\ 
\TT^\alpha_{nt}  &:=   \S^\alpha_{nn} \C^\alpha_{nt} \\
\TT^\alpha_{tt}  &:=  \C^\alpha_{tt}- \C^\alpha_{tn} \S^\alpha_{nn} \C^\alpha_{nt}.
\end{align}

\subsection{Multiphase-field model for partitioning}\label{sec:part}

The consideration that the austenite-martensite $\gamma\alpha'$-interface remains stationary during partitioning obviates the need for solving the phase-field evolution equation~\cite{toji2014atomic,toji2015carbon}. 
Despite this simplification, the partitioning demands a thermodynamically-consistent diffusion of the carbon.
Therefore,  an established model, based on the grand-potential functional, is involved in the present work to exclusively simulate the partitioning of carbon.
With the emphasis to avoid any contribution of the bulk phases to the interface, it has been shown that a more simplistic and a computationally efficient approach can be phenomenologically actualised, if the model is derived from the grand-potential functional with chemical potential as the dynamic variable~\cite{plapp2011unified}.
An extensive description of present model for a multiphase multicomponent system has already been reported, wherein  the sharp-interface solutions are recovered through asymptotic analysis~\cite{choudhury2012grand,amos2018phase}. 
However, in this section, a brief and a contextual derivation of the phase-field model is extended.
Since the current work analyses the evolution in a binary Fe-C system, as opposed to the previous derivations, the concentration and chemical potential are not represented using a vector notation~\cite{amos2018mechanisms,amos2018globularization,amos2018volume}.
Moreover, the notations are further simplified by not indicating their dependence of $T$, owing to the isothermal consideration in the present study.

The grand-potential functional, which governs the partitioning of carbon in a domain of volume $V$, is defined as
\begin{align}\label{eq:GP_functional}
{\Omega}(\mu,\vphi)=\int_{V}\Big[\bar{\Psi}_{\text{bulk}}(\mu,\vphi)+\bar{\Psi}_{\text{intf}}(\vphi, \n \vphi)\Big]\text{d}V,
\end{align}
where $\bar{\Psi}_{\text{bulk}}$ and $\bar{\Psi}_{\text{intf}}$ represent the grand-potential density contribution from the bulk phases and the interfaces, respectively. 
The interface formulation in the elastic model is adopted here, i.e. $\bar{\Psi}_{\text{intf}}(\vphi, \n \vphi) = \bar{W}_\text{intf}(\vphi, \n \vphi)$ in Eqn.~\ref{eq:functional}.

The grand-potential density of the bulk phases are expressed as an interpolation of the grand-potential densities of the individual phases
\begin{align}\label{eq:interpolation_grandchem}
\bar{\Psi}_{\text{bulk}}\left(\mu,\vphi\right) =
\sum_{\alpha=1}^N\Psi_{\alpha}\left(\mu\right)h_{\alpha}\left(\vphi\right),
\end{align}
with $h_{\alpha}\left(\vphi\right)$ and $\mu$ representing the interpolation function and the chemical potential, respectively.
The grand-potential density of the individual phase $\Psi_{\alpha}$ reads
\begin{align}\label{eq:ind_grandchem}
\Psi_{\alpha}\left(\mu\right) =
f_\alpha\left(c^{\alpha}(\mu\right))- \mu c^{\alpha}(\mu),
 \end{align}
where $c^{\alpha}$ corresponds to the carbon concentration in $\alpha$.
The temporal evolution of the concentration, which dictates the partitioning kinetics, is written as
\begin{align}\label{eq:conc_evolution}
\dfrac{\partial \bar{c}}{\partial t}= 
\n \cdot \Big( \sum_{\alpha=1}^N \underbrace{D^{\alpha}(\bar{c})\dfrac{\partial c^{\alpha}\left(\mu\right)}{\partial \mu}}_{:=\bar{M}_{\alpha}} g_{\alpha}(\vphi) \n \mu\Big),
\end{align}
where $g_{\alpha}(\vphi)$ interpolates the mobility $\bar{M_{\alpha}}$~\cite{choudhury2012grand}.
Here, it is important to note that, while $c^{\alpha}$ represents the carbon concentration in $\alpha$, the continuous concentration variable is denoted by $\bar{c}$, which is subsequently introduced in Eqn.~\ref{eq:chem_pot_2}.
Furthermore, the diffusivity $D^{\alpha}(\bar{c})$ in Eqn.~\ref{eq:conc_evolution} not only varies with the phases, but is also influenced by the concentration $\bar{c}$. 
This influence of concentration on diffusivity is discussed later (Sec.~\ref{sec:diff}). 

In this formulation, the Legendre transform of the free energy $f_{\alpha}(\bar{c})$, the grand-potential density $\Psi_{\alpha}$, replaces the conventional dynamic variable $\bar{c}$ with the chemical potential $\mu$ and yields the relation
 \begin{align}\label{eq:conc}
\bar{c} = -\dfrac{\partial \bar{\Psi}_{\text{bulk}}\left(\mu,\vphi\right)}{\partial \mu}.
\end{align}
By incorporating Eqn.~\ref{eq:interpolation_grandchem} in the above relation (Eqn.~\ref{eq:conc}), the concentration reads
 \begin{align}\label{eq:chem_pot_1}
\bar{c} = -\Big( \sum_{\alpha=1}^N \dfrac{\partial \Psi_{\alpha}\left(\mu,\vphi\right)}{\partial \mu} h_{\alpha}\left(\vphi\right)\Big),
\end{align}
which can be further simplified to
 \begin{align}\label{eq:chem_pot_2}
\bar{c} =  \sum_{\alpha=1}^N c^{\alpha}(\mu) h_{\alpha}\left(\vphi\right).
\end{align}
However, from the above Eqn.~\ref{eq:chem_pot_2}, the evolution of the concentration is written as
 \begin{align}\label{eq:chem_pot_3}
\dfrac{\partial \bar{c}}{\partial t}= \dfrac{\partial \bar{c}(\mu,\vphi)}{\partial \mu} \dfrac{\partial \mu}{\partial t}+ \dfrac{\partial \bar{c}(\mu,\vphi)}{\partial \vphi} \dfrac{\partial \vphi}{\partial t},
\end{align}
wherein $\dfrac{\partial \bar{c}(\mu,\vphi)}{\partial \mu}$ and $\dfrac{\partial \bar{c}(\mu,\vphi)}{\partial t}$ can be expressed as
\begin{align}\label{eq:chem_pot_4}
\dfrac{\partial \bar{c}(\mu,\vphi)}{\partial \mu} = \sum_{\alpha=1}^N \dfrac{\partial c^{\alpha}(\mu)}{\partial \mu} h_{\alpha}\left(\vphi\right)
\end{align}
and 
\begin{align}\label{eq:chem_pot_5}
\dfrac{\partial \bar{c}(\mu,\vphi)}{\partial t} = \sum_{\alpha=1}^N  c^{\alpha}(\mu) \dfrac{\partial h_{\alpha}\left(\vphi\right)}{\partial t},  
\end{align}
respectively.
By rearranging Eqn.~\ref{eq:chem_pot_3}, and substituting  Eqns.~\ref{eq:chem_pot_4} and~\ref{eq:chem_pot_5}, the evolution equation for the chemical potential reads
\begin{align}\label{eq:ChemPot_eqn}
\dfrac{\partial \mu}{\partial t} = \Big[\n \cdot \Big( \sum_{\alpha=1}^N D^{\alpha}(\bar{c})\dfrac{\partial c^{\alpha}\left(\mu\right)}{\partial \mu} g_{\alpha}(\vphi) \n \mu\Big) - \sum_{\alpha=1}^N  c^{\alpha}(\mu) \dfrac{\partial h_{\alpha}\left(\vphi\right)}{\partial t} \Big] \Big[ \sum_{\alpha=1}^N \dfrac{\partial c^{\alpha}(\mu)}{\partial \mu} h_{\alpha}\left(\vphi\right) \Big]^{-1}.
\end{align}
As elucidated in this section, two distinct functions, $h_{\alpha}\left(\vphi\right)$ and $g_{\alpha}\left(\vphi\right)$, are conventionally adopted to interpolate the continuous variables across the interface.
However, it has recently been shown that the contribution from the bulk and interface can be efficiently decoupled in a chemo-elastic model by interpolating the variables through the phase field~\cite{amos2018chemo}
In other words, replacing the interpolation functions with the phase field enhances the efficiency of the approach while retaining the thermodynamically consistency.
Therefore, in this work, the concentration and the chemical potential are interpolated through the phase-field variable $\phi_\alpha$.

\subsection{Incorporation of the constrained carbon equilibrium (CCE) condition}\label{sec:incor_cce}

Subscribing to the consideration that the carbide formation and the interface migration is precluded during the partitioning of the Q\&P treatment, a noticeable deviation from the CALPHAD based equilibrium condition is anticipated. 
Particularly, the composition of the constituent phases at the end of the partitioning, which directly governs the amount of retained austenite, is assumed to be consistent with the specific equilibrium condition, referred to as constrained carbon equilibrium (CCE).
Therefore, by imposing the thermodynamical criterion that the chemical potential of the carbon is equal at the end of the partitioning, the respective concentration of the phases is determined.
Correspondingly, equating the activity of carbon yields the relation
\begin{align}\label{eq:cce_1}
 c^{\gamma}=c^{\alpha'} \cdot \exp\Big(\frac{76789-43.8T-(169105-120.4T) \cdot c^{\gamma}}{\kappa T}\Big),
\end{align}
where $c^{\gamma}$ and $c^{\alpha'}$ represent the respective mole fraction of carbon in austenite and martensite, which is employed for the calculation of the endpoints of the partitioning~\cite{lobo1976thermodynamics,lobo1976thermodynamics2}.

The stationary nature of the interface indicates that carbon exclusively diffuses across the interface during the partitioning.
Therefore, the number of iron atoms are conserved across the phases, which can be expressed
\begin{align}\label{eq:cce_2}
 f^{\gamma}_{\text{CCE}}(1-c^{\gamma}_{\text{CCE}})=f^{\gamma}_{i}(1-c^{\text{alloy}}),
\end{align}
where $c^{\gamma}_{\text{CCE}}$ and $c^{\text{alloy}}$ represent the mole fraction of carbon after partitioning in austenite and the alloy composition, respectively.
Furthermore, while $f^{\gamma}_{\text{CCE}}$ denotes the mole fraction of austenite after partitioning, the initial mole fraction of austenite is given by $f^{\gamma}_{i}$.
Although the inequality between $f^{\gamma}_{\text{CCE}}$ and $f^{\gamma}_{i}$ apparently contradicts the assumption of the stationary interface, the negligible change in the volume fraction of the phases, during the extensive diffusion of carbon, accounts for the disparity. 
From Eqns.~\ref{eq:cce_1},~\ref{eq:cce_2} and the mass balance relations, expressed as
\begin{align}\label{eq:cce_3}
 f^{\alpha'}_{\text{CCE}}c^{\alpha'}_{\text{CCE}}+f^{\gamma}_{\text{CCE}}c^{\gamma}_{\text{CCE}}=c^{\text{alloy}},
\end{align}
and
\begin{align}\label{eq:cce_4}
 f^{\alpha'}_{\text{CCE}}+f^{\gamma}_{\text{CCE}}=1,
\end{align}
the parameters $c^{\alpha'}_{\text{CCE}}$, $c^{\gamma}_{\text{CCE}}$, $f^{\gamma}_{\text{CCE}}$ and $f^{\alpha'}_{\text{CCE}}$, which characterize CCE condition, are determined.

As opposed to the conventional incorporation of the CALPHAD data~\cite{amos2018phase}, the introduction of the CCE condition demands a certain degree of manipulation.
To this end, the martensite is treated as supersaturated ferrite.
Moreover, it has been conceded that the CCE condition focuses extensively on establishing the endpoints of the partitioning~\cite{speer2003carbon}.
Therefore, the numerical manipulation, henceforth presented, is primarily aimed at establishing the characteristic CCE parameters, while being tangible with the CALPHAD data for the chemical driving-force.

In the binary Fe-C system, owing to the interstitial occupation of the carbon atoms, the free energy of a given phase $\Theta$, which can be $\gamma$ or $\alpha'$, is expressed as
\begin{align}\label{eq:interstitial}
 f^{\Theta}(Y(c))=Y^{2}_{Va}f_{Fe:Va}+  Y^{2}_{C}f_{Fe:C}+RT[\tilde{a^2}(Y^2_{C}\ln Y^2_{C}
  +Y^2_{Va}\ln Y^2_{Va})]+Y^2_{C}Y^2_{Va}\bar{L}_{Fe:Va,C}.
\end{align}
In this formulation, the dependence of the free energy on the concentration is expressed in terms of the site fraction $Y_i^s$, where $s$ denotes the sub-lattices ($1$ or $2$) and $i$ represents the component occupying the sub-lattice.
The sub-lattice 1 is occupied by iron ($Fe$) and a fraction of the sub-lattice $2$ is occupied by carbon ($C$), based on the carbon concentration, leaving the remaining sites vacant ($Va$). 
The site-fraction is related to the mole fraction through the expression $c_i=\frac{{\underset{s}{\sum}}a^sY_i^s}{{\underset{s}{\sum}}a^s(1-Y_{Va}^s)}$, where $a^s$ is the number of sites in the sub-lattice $s$ per unit mole. 
Moreover, $f_{Fe:Va}$ and $f_{Fe:C}$ correspond to the free energy of pure iron and Fe-C steel, respectively. 
The constant $\tilde{a^2}$ varies with the crystal structure of the phases. For ferrite $\tilde{a^2} = 3$, while for austenite $\tilde{a^2} = 1$. $\bar{L}_{Fe:Va,C}$ denotes the interaction parameter with comma and colon separating the components and the sub-lattices, respectively. 
The extensive formulation of the free energy in Eqn.~\ref{eq:interstitial} can be simplified using mole fraction~\cite{amos2018chemo} and written as
\begin{align}\label{eq:cce_5}
 f^{\Theta}(c) = A^{\Theta}c^2+B^{\Theta}c+D^{\Theta}.
\end{align}
Correspondingly, the first and second derivatives of the free energy are expressed as 
\begin{align}\label{eq:cce_6}
 \frac{\partial f^\Theta}{\partial c}=2A^{\Theta}c+B^{\Theta}=\frac{1}{V_\text{m}}\cdot\frac{\partial G^\Theta}{\partial c}=\mu^{\Theta}
\end{align}
and
\begin{align}\label{eq:cce_7}
 \frac{\partial^2 f^\Theta}{\partial c^2}=2A^{\Theta}=\frac{1}{V_\text{m}}\cdot\frac{\partial^2 G^\Theta}{\partial c^2}=\frac{\partial \mu^{\Theta}}{\partial c},
\end{align}
where $V_\text{m}$ corresponds to the molar volume of the phases and $G^\Theta$ is the Gibbs free-energy which the CALPHAD database renders~\cite{gustafson1985thermodynamic}. 
Though, the numerical co-coefficients $A^{\Theta}$,  $B^{\Theta}$ and $D^{\Theta}$ do not posses any thermodynamic significance, a proper fitting facilitates the replication of the CALPHAD based free-energy plot through these parameters~\cite{amos2018chemo}.
Furthermore, by appropriately manipulating these co-coefficients the CCE condition can be introduced into the simulation. 

The numerical approach, which relates the co-coefficients with the CALPHAD data, begins by defining a concentration-step of finite magnitude, $\Delta c$~\cite{amos2018phase}.
The Gibbs free-energy at a specific composition, which is considered to be the equilibrium composition ($c_{\text{eq}}$) here, and the neighbouring compositions defined by $\Delta c$ are determined from the database. 
Thus, the free energies $G^{\Theta}_{c_{\text{eq}}}$, $G^{\Theta}_{c_{\text{eq}}^{1}}$, $G^{\Theta}_{c_{\text{eq}}^{2}}$, $G^{\Theta}_{\tilde{{c_{\text{eq}}^{1}}}}$ and $G^{\Theta}_{\tilde{{c_{\text{eq}}^{2}}}}$ correspond to the composition $c_{\text{eq}}$, $c_{\text{eq}}+\Delta c$, $c_{\text{eq}}+2\Delta c$, $c_{\text{eq}}-\Delta c$ and $c_{\text{eq}}-2\Delta c$, respectively.
An \lq environment \rq \thinspace surrounding the equilibrium composition is re-created around the predetermined $c_{\text{CCE}}$ through the following consideration
\begin{align}\label{eq:cce_11}
 G^{\Theta}_{c_{\text{CCE}}^1} = G^{\Theta}_{c_{\text{CCE}}} + (G^{\Theta}_{c_{\text{eq}}^1} - G^{\Theta}_{c_{\text{eq}}}),
\end{align}
\begin{align}\label{eq:cce_12}
 G^{\Theta}_{c_{\text{CCE}}^2} = G^{\Theta}_{c_{\text{CCE}}} + (G^{\Theta}_{c_{\text{eq}}^2} - G^{\Theta}_{c_{\text{eq}}}),
\end{align}
\begin{align}\label{eq:cce_13}
 G^{\Theta}_{\tilde{{c_{\text{CCE}}^1}}} = G^{\Theta}_{c_{\text{CCE}}} + (G^{\Theta}_{\tilde{{c_{\text{eq}}^1}}} - G^{\Theta}_{c_{\text{eq}}}),
\end{align}
and 
\begin{align}\label{eq:cce_14}
 G^{\Theta}_{\tilde{{c_{\text{CCE}}^2}}} = G^{\Theta}_{c_{\text{CCE}}} + (G^{\Theta}_{\tilde{{c_{\text{eq}}^2}}} - G^{\Theta}_{c_{\text{eq}}}),
\end{align}
wherein $G^{\Theta}_{c_{\text{CCE}}}$, the free energy of the phases at $c^{\Theta}_{\text{CCE}}$ is obtained from the CALPHAD database. 

If $A^{\Theta}_{\text{CCE}}$, $B^{\Theta}_{\text{CCE}}$ and $D^{\Theta}_{\text{CCE}}$ are the modified co-coefficients which introduce the CCE conditions, the adapted free energy is represented by
\begin{align}\label{eq:cce_9}
 f^{\Theta}_{\text{CCE}}(c) = A^{\Theta}_{\text{CCE}}c^2+B^{\Theta}_{\text{CCE}}c+D^{\Theta}_{\text{CCE}} = \frac{1}{V_\text{m}}G^{\Theta}_{\text{CCE}}(c).
\end{align}
By substituting Eqns~\ref{eq:cce_11}-\ref{eq:cce_14} and adopting a five-point stencil scheme in finite-difference approximation~\cite{amos2018phase,amos2018chemo}, the co-coefficients are determined by 
\begin{align}\label{eq:coeff_1}
  \frac{\partial^2G^{\Theta}_{\text{CCE}}}{\partial c^2}=\frac{-G^{\Theta}_{c_{\text{CCE}}^2}+16G^{\Theta}_{c_{\text{CCE}}^1}-30G^{\Theta}_{c_{\text{CCE}}}+16G^{\Theta}_{\tilde{{c_{\text{CCE}}^1}}}-G^{\Theta}_{\tilde{{c_{\text{CCE}}^2}}}}{12(\Delta c)^2}=2A^{\Theta}_{\text{CCE}}
\end{align}
and
\begin{align}\label{eq:coeff_2}
  \frac{\partial G^{\Theta}_{\text{CCE}}}{\partial c}=\frac{-G^{\Theta}_{c_{\text{CCE}}^2}+8G^{\Theta}_{c_{\text{CCE}}^1}-8G^{\Theta}_{\tilde{{c_{\text{CCE}}^1}}}+G^{\Theta}_{\tilde{{c_{\text{CCE}}^2}}}}{12\Delta c}=2A^{\Theta}_{\text{CCE}}c+B^{\Theta}_{\text{CCE}}.
\end{align}
Furthermore, the condition
\begin{align}\label{eq:cce_10}
 2A^{\alpha'}_{\text{CCE}}c^{\alpha'}_{\text{CCE}}+B^{\alpha'}_{\text{CCE}}=2A^{\gamma}_{\text{CCE}}c^{\gamma}_{\text{CCE}}+B^{\gamma}_{\text{CCE}} = \frac{f^{\gamma}_{\text{CCE}}-f^{\alpha'}_{\text{CCE}}}{c^{\gamma}_{\text{CCE}}-c^{\alpha'}_{\text{CCE}}} 
\end{align}
is invoked in the \emph{Newton-Raphson iteration} technique, to ensure that the modified free-energy formulation (Eqn.~\ref{eq:cce_9}) yields equal chemical potential at the compositions $c^{\gamma}_{\text{CCE}}$ and $c^{\alpha'}_{\text{CCE}}$. 
The thermodynamic data for the stoichiometric compounds are introduced by numerically fitting a sharp curve with its minima at the point rendered by the CALPHAD data~\cite{amos2018phase}. 
Although the present approach is motivated by this methodology of introducing the stoichiometric compounds, the $\Delta c$ is appropriately chosen such that   
through Eqns~\ref{eq:cce_11}-\ref{eq:cce_14} a tangibility with the CALPHAD data is attained.

\subsection{Carbon diffusivity}\label{sec:diff}

The kinetics of the carbon diffusion, during the partitioning, is governed by its diffusivity in the corresponding phases. 
To encompass the influence of the carbon concentration on the diffusivity, the following formulations,
\begin{align}\label{eq:d_gamma}
 D^{\gamma}(c) = 4.53 \times 10^{-7} \Big( 1+y_{c}(1-y_{c})\frac{8339.9}{T} \Big)\cdot\exp\Big[-(\frac{1}{T}-2.21 \times 10^{-4})(17767-26436y_{c}) \Big]
\end{align}
and 
\begin{align}\label{eq:d_alpha}
 D^{\alpha'}= 0.02 \times 10^{-4}\cdot\exp\Big(\frac{-10115}{T}\Big)\cdot\exp\Bigg\{ 0.5898\Big[1+\frac{2}{\pi}\text{arctan}\Big(1.4985-\frac{15309}{T}\Big)\Big]\Bigg\}
\end{align}
are incorporated in the model to account for the diffusion constant in austenite ($D^{\gamma}(c)$) and martensite ($D^{\alpha'}$), respectively~\cite{aagren1982computer,aagren1986revised,seo2016kinetics}. 
In Eqn.~\ref{eq:d_gamma}, the carbon concentration is incorporated as $y_c=c/(1-c)$.
Owing to the difference in the diffusivities, it is shown that the carbon accumulates in the $\gamma\alpha'$-interface during the partitioning~\cite{toji2015carbon,takahama2012phase}. 
From Eqn.~\ref{eq:d_gamma}, it is conceivable that the accumulation of the carbon, particularly under CCE conditions, increases the diffusivity in austenite beyond $D^{\alpha'}$. To preclude such nonphysical changes in the diffusivities, a cut-off is defined, which ensures that a minimum difference in the diffusion constants is retained at $10^{-3}$ at any stage of the transformation~\cite{mecozzi2016phase}.

\subsection{Domain set-up}

The 2-dimensional domain, considered in the present work, is spatially discretised through the finite-difference approach. 
This discretisation decomposes the domain into uniform grids of dimension $\Delta\text{x}=\Delta\text{y}=0.05~\upmu\text{m}$. 
The domain size of dimension 800 $\times$ 800 cells, rendering 40~$\upmu$m $\times$ 40~$\upmu$m domain, is considered for all simulations. 
The evolution equations formulated in Sec.~\ref{sec:quench} and Sec.~\ref{sec:part} are solved on the grid points using finite-difference algorithm with an explicit forward Euler scheme. 
The length parameter $\epsilon$, which dictates the width of the diffuse interface, is fixed at $2.5~\Delta\text{x}$~\cite{choudhury2012grand}.

The numerical efficiency of the present approach is enhanced by incorporating the models to the in-house software package \pace{} (Parallel Algorithms for Crystal Evolution in 3D).
Furthermore, the consumption of the computational resources is optimised through the domain decomposition using Message Passing Interface (MPI).
The polycrystalline set-up is achieved through the Voronoi tessellation which yields 200 grains~\cite{aurenhammer1991voronoi,perumal2018phase}. 
In order to achieve a time-invariant normal distribution of the grains, which are devoid of any artifacts, the microstructure is allowed to evolve in an isotropic condition, till the total number of grains is reduced to 100.

\section{Results and discussion}

\subsection{Displacive martensite transformation}

The volume fraction of martensite ($f^{\alpha'}$), which results from quenching, influences the amount of retained austenite both directly and indirectly. 
In a two phase system of martensite and austenite, $f^{\alpha'}$ ascertains the amount of austenite available for any transformations.
Furthermore, since the endpoints of the partitioning vary with the martensite volume-fraction, Eqns.~\ref{eq:cce_1}-~\ref{eq:cce_4}, $f^{\alpha'}$ indirectly influences the amount of retained austenite.

To simulate a definite volume fraction of martensite, an Fe-C system with carbon concentration $c=0.009$ ($\text{C}=0.2~\text{wt}\%$) is considered. 
The quenching temperature which yields the desired martensite volume-fraction is calculated from the Koistinen-Marburger equation~\cite{koistinen1959general}.
This relation between the volume fraction $f^{\alpha'}$ and temperature ($T$) is expressed as
\begin{align}\label{eq:KM_Eqn}
 f^{\alpha'}=1-\exp[-\alpha_{m}(\Ms-T)],
\end{align}
where $\Ms$ is the martensite start temperature. 
The $\Ms$ temperature varies with the carbon content. 
From Ref.~\cite{van2012bainite}, $\Ms$ temperature for a given concentration is calculated by
\begin{align}\label{eq:ms_temp}
\Ms(\text{C})=565-600[1-\exp(-0.96\text{C})],
\end{align}
where $\alpha_{m}$, in Eqn.~\ref{eq:KM_Eqn}, is the rate parameter which is written as
\begin{align}\label{eq:rate_constant}
\alpha_{m}=27.2-19.8[1-\exp(-1.56\text{C})].
\end{align}
In the aforementioned Eqns.~\ref{eq:KM_Eqn}-~\ref{eq:rate_constant}, the concentration $\text{C}$ is the weight percent of carbon, while the temperatures ($T$ and $\Ms$) are considered in $\degree \text{C}$. 
Three different volume fractions of martensite are arbitrarily chosen: $f^{\alpha'}$= 0.4, 0.6 and 0.8, and the corresponding quenching temperatures are determined from Eqn.~\ref{eq:KM_Eqn}.
The chemical driving-forces pertaining to the quenching temperatures, which are deduced from the CALPHAD database, are presented in Table~\ref{tab:chem_drive}.
\begin{table}[!ht]
\centering
\caption{Chemical driving forces at different quenching temperatures ascertained from CALPHAD database and prefactor}
\scalebox{0.9}{
\begin{tabular}{l|l|l} 
\hline
{ $T$ ($\degree \text{C}$)} & { $\Delta W_{\text{chem}}^{\alpha'}$ (J/m$^3$)} & {$f_{\text{c}}$}  \\ \hline	
437	& $- 2.2594 \times 10^8$  	&  0.78\\ \hline
418	& $- 2.4509 \times 10^8$  	&  0.695\\ \hline
386	& $- 2.7814 \times 10^8$  	& 0.66\\ \hline
\end{tabular}
}
\label{tab:chem_drive}
\end{table}

The diffusionless transformation of martensite is accompanied by the change in the crystal structure from Face-Centered Cubic (FCC-austenite) to Body-Centered Tetragonal (BCT-martensite).
The transition in the crystal structure introduces strain owing to the difference in the lattice parameters.
This strain, which is induced despite the absence of any external stresses, is conventionally referred to as stress-free transformation strain or eigenstrain.
In the present work, the eigenstrains are introduced as phase-dependent Bain strains which are written as
\begin{align}\label{eq:Bain_Strains}
\tilde{\vv{\epsilon}}^{000}(1) =
\begin{pmatrix}
 \epsilon_3 &          0 &          0 \\
          0 & \epsilon_1 &          0 \\
          0 &          0 & 0 \\
\end{pmatrix}
, \quad \tilde{\vv{\epsilon}}^{000}(2)=
\begin{pmatrix}
 \epsilon_1 &          0 &          0 \\
          0 & \epsilon_3 &          0 \\
          0 &          0 &          0 \\
\end{pmatrix}.
\end{align}
The eigenstrains $\epsilon_1$ and $\epsilon_3$ which dictate the Bain strains are calculated from the crystal lattice parameter.
The formulation used for determining the lattice parameter and its other implications in the present work are consolidated in the Appendix.
Since the grains in a polycrystalline microstructure are rotated randomly, the eigenstrains are correspondingly rotated through the rotation matrix $Q$, in the $z-y'-x''$ intrinsic rotation convention.
Therefore, the elastic strains become $\tilde{\epsilon}^{00}_{ij} =  Q_{im} Q_{jn} \tilde{\epsilon}^{000}_{mn}$, depending on the crystal orientation of the grain.
Furthermore, the misfit strain relaxation at austenite grain boundaries~\cite{heo2014phase} is included by 
\begin{align}\label{eq:misfit}
 \Gamma(\vv\phi^{t_0}) =  1 - \rho 16 \sum_{\alpha < \beta} (\phia^{t_0})^2(\phib^{t_0})^2
\end{align}
where the factor $0 \leq \rho \leq 1$ determines the strength of the relaxation.
In Eqn.~\ref{eq:misfit}, $\phia^{t_0}$ and $\phib^{t_0}$ are the phase-field variables which are assigned once the diffuse interface is established between the grains (at time $t=0$).
The misfit strain relaxation is accommodated in the eigenstrain matrix by $\tilde{\vv{\epsilon}}^{0} =  \Gamma(\vv\phi^{t_0})\tilde{\vv{\epsilon}}^{00}$.

\begin{figure}
\centering
   \begin{subfigure}[b]{0.85\textwidth}
   \includegraphics[width=1\linewidth]{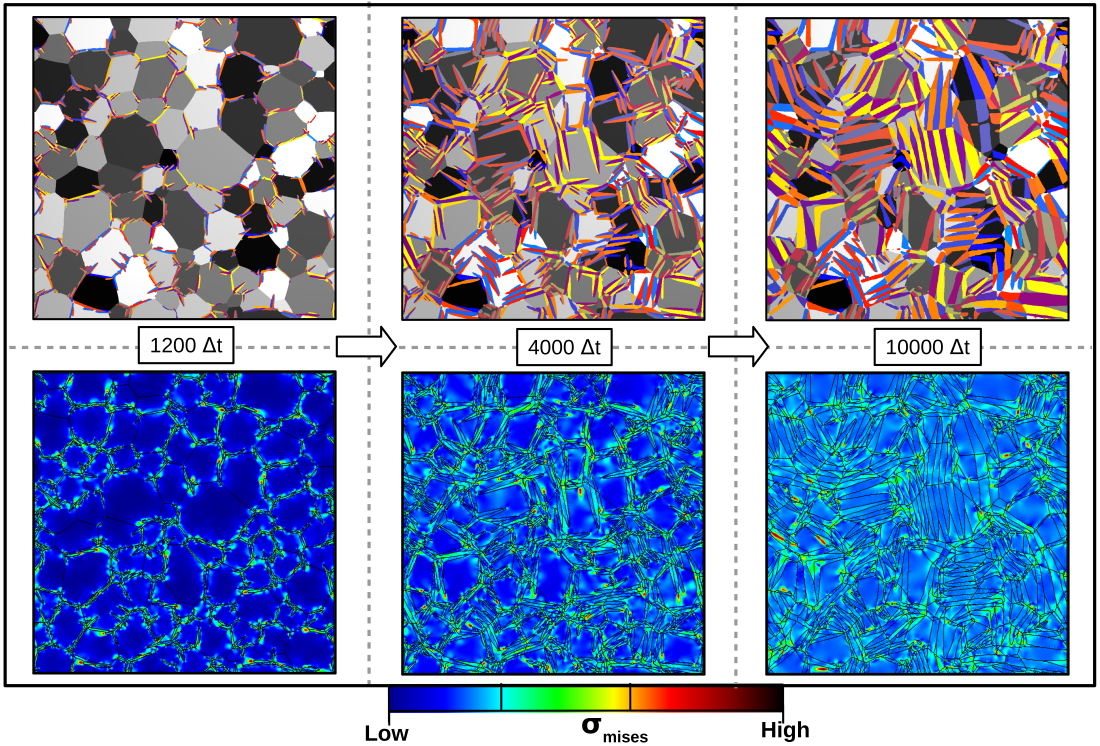}
   \caption{}
   \label{fig:M_evolution} 
\end{subfigure}

\begin{subfigure}[b]{0.85\textwidth}
   \includegraphics[width=1\linewidth]{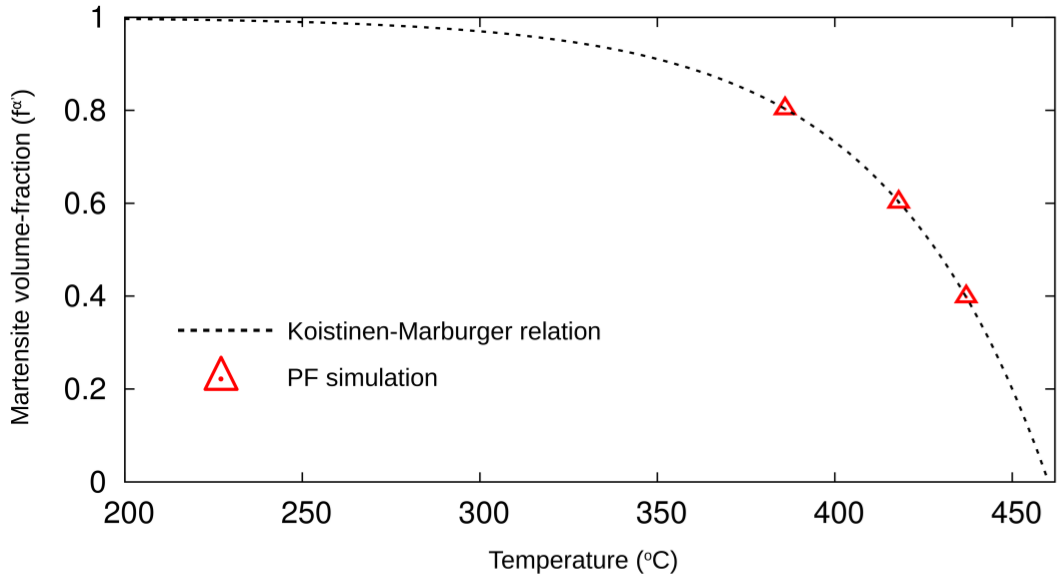}
   \caption{}
   \label{fig:KM_plot}
\end{subfigure}

\caption[Two numerical solutions]{a) The diffusionless decomposition of austenite to martensite at $418~\degree \text{C}$ in an Fe-C alloy of $0.2~\text{wt}\%$ carbon. The evolution of the von Mises stress ($\sigma_\text{mises}$) indicates the elastic strain accompanying the martensite transformation. b) The volume fraction of the martensite resulting from the phase-field simulation (PF simulation) at three selected temperatures is plotted along with the Koistinen-Marburger relation. }
\end{figure}

Although the Bain strains, which govern the martensite evolution, account for the elastic strains, experimental investigations indicate that the displacive transformation is additionally accompanied by plastic strains~\cite{krauss1999martensite}.
Since plasticity is not included in the current phase-field model, the components of the Bain strains are reduced by a prefactor $f_{\text{c}}$, which transforms the eigenstrains to  $\epsilon_1 = \epsilon_1^0 f_{\text{c}}$ and $\epsilon_3 = \epsilon_3^0 f_{\text{c}}$.
The prefactor avoids unphysically high elastic driving-forces~\cite{hueter2015multiscale}.
The different prefactors, $f_{\text{c}}$, which vary with the temperature, are presented in Table~\ref{tab:chem_drive}.
The material parameters, encompassing both elastic and interface parameter, which are incorporated in the model are tabulated in Table~\ref{tab:parameter_elast}.
\begin{table}[!ht]
\begin{center}
\caption{Parameter for the quenching simulation}
\begin{tabular*}{0.7\textwidth}{l c c} \hline
Parameter      				& Symbol 			& Value  			\\ \hline
Young's modulus  			& $E$           		& 200 GPa 			\\ \hline	
Poisson's ratio			& $\nu$ 			& 0.3				\\ \hline	
Eigenstrain 				& $\epsilon_1^0$ 		& 0.14				\\ \hline	
Eigenstrain  				& $\epsilon_3^0$ 		& -0.08 			\\ \hline
Interfacial energy  			& $\gab$ 			& 0.5 J/m$^2$			\\ \hline	
Interface parameter	  		& $\gamma_\alpha^c$		& 3.0 J/m$^2$			\\ \hline
Third order parameter  			& $\gabd$ 			& 2 ($\gab + \gamma_\alpha^c$)	\\ \hline
Misfit strain relaxation parameter	& $\rho$ 			& 0.05				\\ \hline	
\end{tabular*} 
\label{tab:parameter_elast}
\end{center}
\end{table}
Since the martensite exhibits a rapid transformation rate, a refined time-step $\Delta t$ with no physical significance is considered, despite the consideration of the quantitative driving force.
For all simulations, the noise term in Eq.~\ref{eq:phase_field_evolution} is applied every hundredth time step with a uniform distribution which is scaled to result in a fluctuation up to 0.2 in the phase fields.
In the first time step, a three-fold higher noise is employed to stimulate first-time nucleation.
It should be pointed out a noise of phase $\alpha$ is only activated in its respective parental grain.

The microstructural changes accompanying the diffusionless transformation of austenite in a polycrystalline set-up are presented in Fig.~\ref{fig:M_evolution}.
Furthermore, the corresponding evolution of the von Mises stress ($\sigma_\text{mises}$), which signifies the accumulation of the elastic energy, is included.
In the initial stages of the transformation, at $t = 1200\Delta t$, it is evident that the mechanical strains are established during the displacive decomposition of austenite.
However, since the chemical driving-force at these early stages are exceedingly high, compared to the mechanical forces, the martensite grows despite the build-up of the stress.
As shown in Fig.~\ref{fig:M_evolution} at $t = 4000\Delta t$, with an increase in the volume fraction (size) of the martensite, the elastic strains increase visibly.
Consequently, the elastic energy becomes comparable to the chemical driving-force and decreases the rate of the transformation considerably.
When the mechanical stress, developed during the diffusionless decomposition, matches the chemical driving force, the transformation halts ($t = 10000\Delta t$).
This balance between the chemical and mechanical (elastic) forces yields a definite volume fraction of martensite.
The amount of martensite emerging from the displacive transformation of the austenite is ascertained for the different temperatures considered in Table~\ref{tab:chem_drive}.
The martensite volume-fractions are plotted along with the Koistinen-Marburger equation in Fig.~\ref{fig:KM_plot}.
This representation reveals a noticeable agreement between the fraction of the decomposed austenite and the analytical prediction.

\subsection{Partitioning}

\subsubsection{Low carbon concentration: \textnormal{Fe}-$0.2~\textnormal{wt}\%$\textnormal{C} ($c^{\textnormal{alloy}}=0.009$)}\label{sec:low_c}

From Eqns.~\ref{eq:cce_1}-~\ref{eq:cce_4}, the endpoints of the partitioning for the different martensite volume-fractions considered in the present work are calculated. 
The CCE composition of the phases are given in Table~\ref{table:CCE}.
\begin{table}[!ht]
\centering
\caption{CCE composition of martensite and austenite in an Fe-C system of $0.009$ mole frac. carbon.} 
\scalebox{0.9}{
\begin{tabular}{l|l|l} 
\hline
{ $T$ ($\degree \text{C}$)} & { $c^{\gamma}_\text{CCE}$ (mole frac.)} & { $c^{\alpha '}_\text{CCE}$ (mole frac.)}  \\ \hline	
437	& $1.42 \times 10^{-2}$  	&  $7.68 \times 10^{-6}$\\ \hline
418	& $2.12 \times 10^{-2}$  	&  $8.84 \times 10^{-6}$\\ \hline
386	& $3.67 \times 10^{-2}$  	&  $1.20 \times 10^{-5}$\\ \hline
\end{tabular}
}
\label{table:CCE}
\end{table}
By reconstructing the free-energy curve, as elucidated in Sec.~\ref{sec:incor_cce}, and adopting the model, formulated in Sec.~\ref{sec:part}, the partitioning of the carbon is simulated.

\begin{figure}
\centering
   \begin{subfigure}[b]{0.85\textwidth}
   \includegraphics[width=1\linewidth]{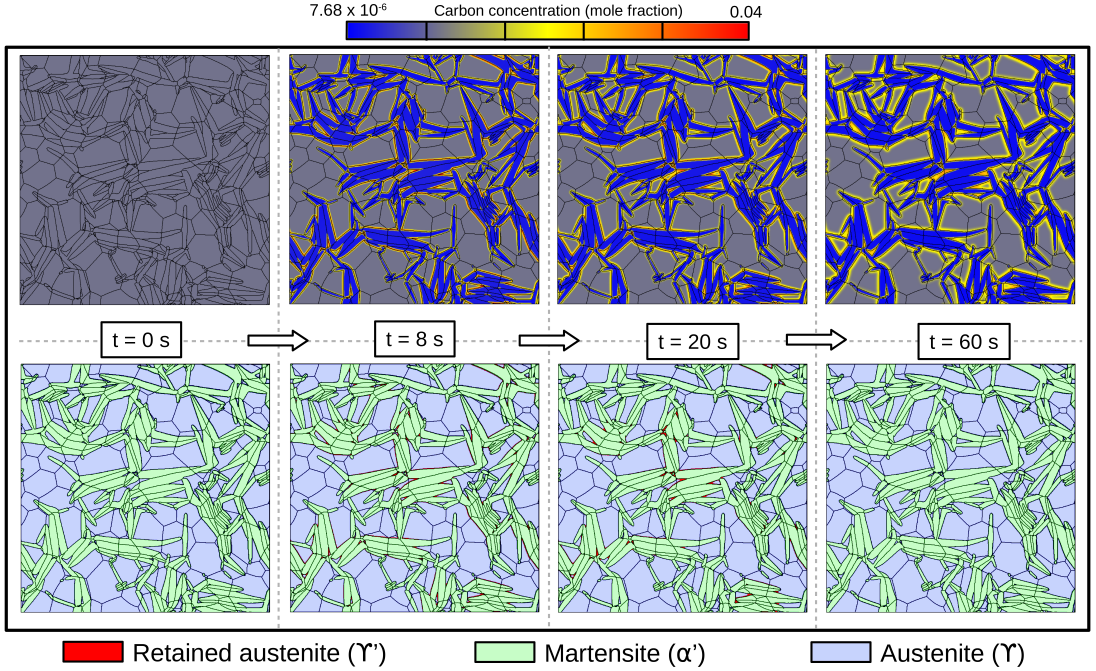}
   \caption{}
   \label{fig:c009v40} 
\end{subfigure}

\begin{subfigure}[b]{0.85\textwidth}
   \includegraphics[width=1\linewidth]{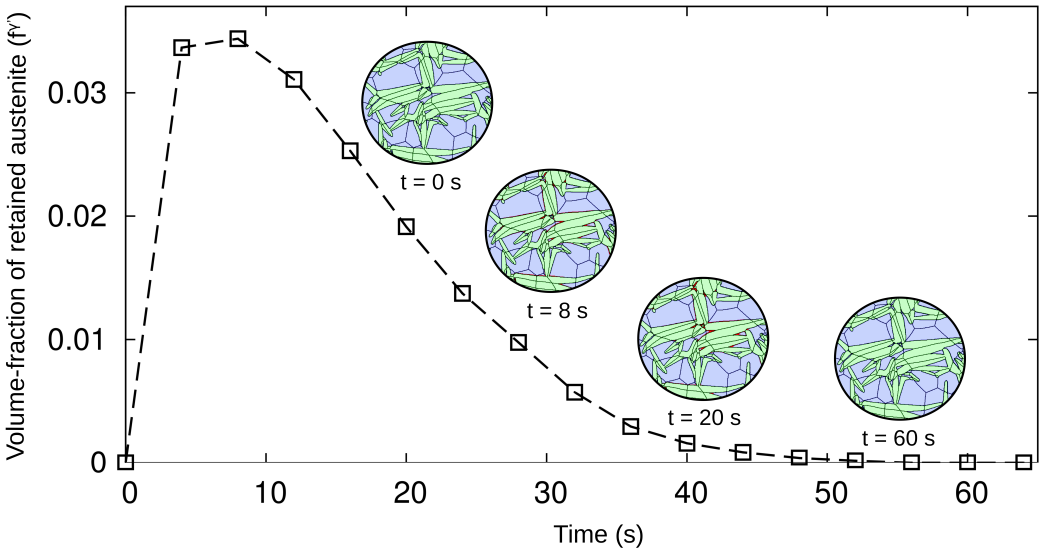}
   \caption{}
   \label{fig:009C_40f}
\end{subfigure}

\caption[Two numerical solutions]{a) The evolution of the carbon concentration during the partitioning at $700~\degree \text{C}$. The potential retained austenite based on the concentration distribution is highlighted.  b) The temporal change in the volume fraction of the retained austenite.}
\end{figure}

Fig.~\ref{fig:c009v40} illustrates the concentration evolution accompanying the partitioning stage. 
From Eqn.~\ref{eq:ms_temp}, the composition which yields stable or retained austenite at room temperature ($25~\degree \text{C}$) is determined. 
Accordingly, from the concentration distribution, the potential retained-austenite is identified and presented in Fig.~\ref{fig:009C_40f}.
Since the primary focus of Q\&P processing is to yield retained austenite, the time-steps chosen to represent the partitioning are confined to the carbon diffusion in austenite, which stabilises $\gamma$.

Owing to the displacive nature of the martensite transformation, the composition remains uniform and unperturbed during quenching ($t=0$~s).
However, during partitioning, the carbon is expelled from martensite to the austenite.
As discussed in Sec.~\ref{sec:diff}, the significant difference in the diffusivities facilitates a faster expulsion of carbon from martensite when compared to its diffusion in austenite.
Therefore, the carbon concentration in the martensite is close to $c^{\alpha '}_\text{CCE}$ from the outset. 
On the other hand, the crystal structure of the corresponding phases, which dictates the rate of carbon diffusion, favours the accumulation of carbon in the austenite grains adjacent to the $\alpha'\gamma$-interface. 
The accumulation of carbon near the interface stabilises the respective austenite at $t=8$~s, as shown in Fig.~\ref{fig:c009v40} and yields potential retained-austenite ($\gamma'$).
But since the accumulated carbon eventually diffuses into the austenite grain, the amount of retained austenite temporally decreases ($t=20$~s) and ultimately vanishes at $t=60$~s.

The change in the volume fraction of the retained austenite ($f^{\gamma'}$), with the progress of partitioning, is illustrated in Fig.~\ref{fig:009C_40f}.
Evidently, the almost immediate increase in the retained austenite lies in contrast to its gradual disappearance. 
In the early stages of the partitioning, the high carbon diffusivity in martensite, which yields the concentration accumulation, is responsible for the drastic increase of $f^{\gamma'}$. 
The subsequent diffusion of carbon within austenite, albeit at a relatively low rate, governs the eventual decrease in retained austenite.
This trend in the temporal evolution of the retained austenite is consistent with the existing report~\cite{mecozzi2016phase}.
Furthermore, the influence of the partitioning duration on the retained austenite, illustrated in Fig.~\ref{fig:009C_40f}, can be employed to define an appropriate thermal cycle that yields stable austenite even in low carbon concentration. 

\begin{figure}
\centering
   \begin{subfigure}[b]{0.85\textwidth}
   \includegraphics[width=1\linewidth]{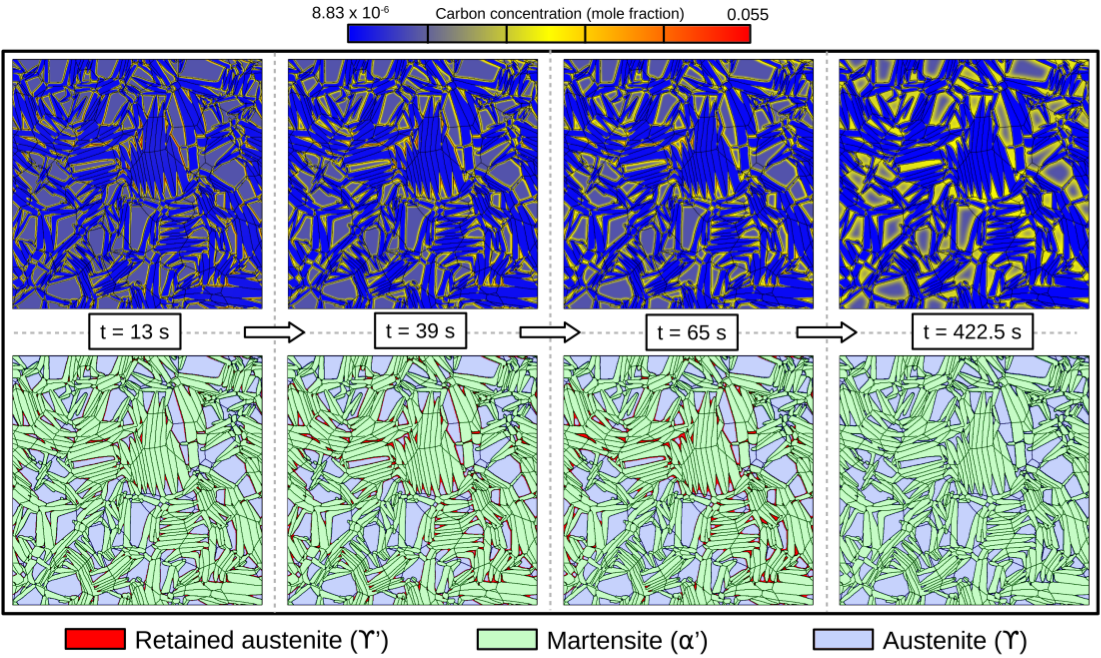}
   \caption{}
   \label{fig:009_60Evolution} 
\end{subfigure}

\begin{subfigure}[b]{0.85\textwidth}
   \includegraphics[width=1\linewidth]{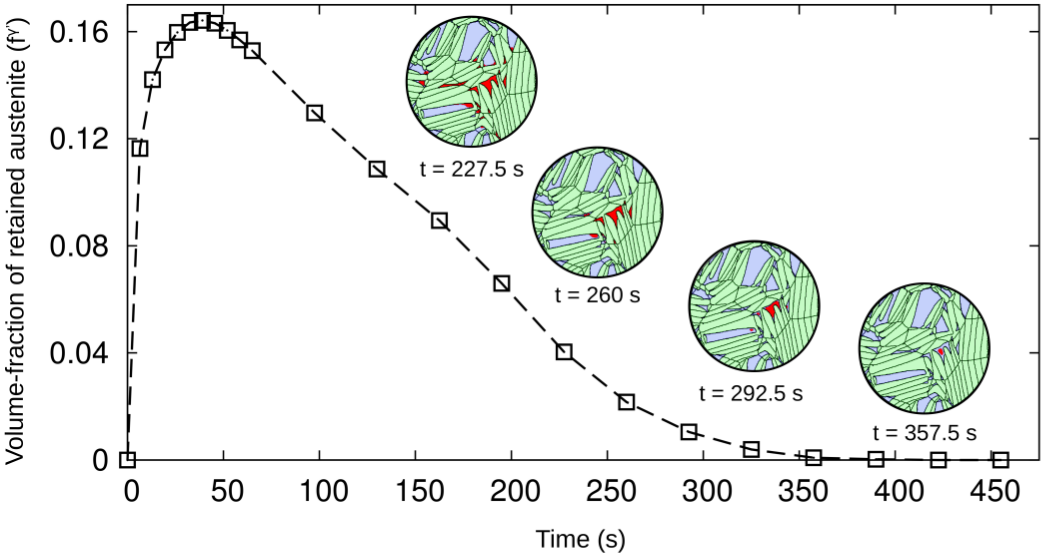}
   \caption{}
   \label{fig:c009_v60}
\end{subfigure}

\caption[Two numerical solutions]{a) The change in the concentration distribution during the partitioning at $700~\degree \text{C}$ in polycrystalline set-up quenched at $418~\degree \text{C}$ ($f^{\alpha'}=0.6$). The formation and subsequent disappearance of retained austenite adjacent to the interface and, more importantly, in austenite islands. b) The evolution of the retained austenite volume-fraction with the prolongation of partitioning.}
\end{figure}

Fig.~\ref{fig:009_60Evolution} shows the partitioning of carbon in the microstructure with increased volume fraction of martensite ($f^{\alpha'}=0.6$).
Similar to Fig.~\ref{fig:c009v40}, the potential retained-austenite, which is ascertained from the carbon distribution, is included.
Despite the slight change in $c^{\alpha '}_\text{CCE}$, the amount of carbon migrating to the austenite increases with increase in the volume fraction of the martensite.
Through accumulation, this carbon expulsion increases the amount of the stable austenite that is formed at the early stages of the partitioning.
Fig.~\ref{fig:c009_v60}, which shows the change in the volume fraction of retained austenite during partitioning, affirms the increased amount of $f^{\gamma'}$ in the initial stages of the transformation.
Moreover, in addition to the drastic increase in $f^{\gamma'}$, which is also observed in Fig.~\ref{fig:009C_40f}, the peak $f^{\gamma'}$ is achieved by the gradual growth of the retained austenite.
This transition from the sudden raise in $f^{\gamma'}$, resulting from the carbon diffusion out of the martensite, to the gradual increase induced by the diffusion within the austenite grains is evident in the early time-steps presented in Fig.~\ref{fig:c009_v60}.
Furthermore, the $f^{\gamma'}$ plot in Fig.~\ref{fig:c009_v60} unravels that, when compared to Fig.~\ref{fig:009C_40f}, the eventual disappearance of the retained austenite is more prolonged in this microstructure.
Although these variations in the concentration distribution and $f^{\gamma'}$ evolution between the $f^{\alpha'}=0.4$ and $f^{\alpha'}=0.6$ microstructures can be described as a natural consequence of the change in the martensite volume-fraction, a closer examination reveals the  significant role of the phase distributions. 

\begin{figure}
    \centering
      \begin{tabular}{@{}c@{}}
      \includegraphics[width=0.8\textwidth]{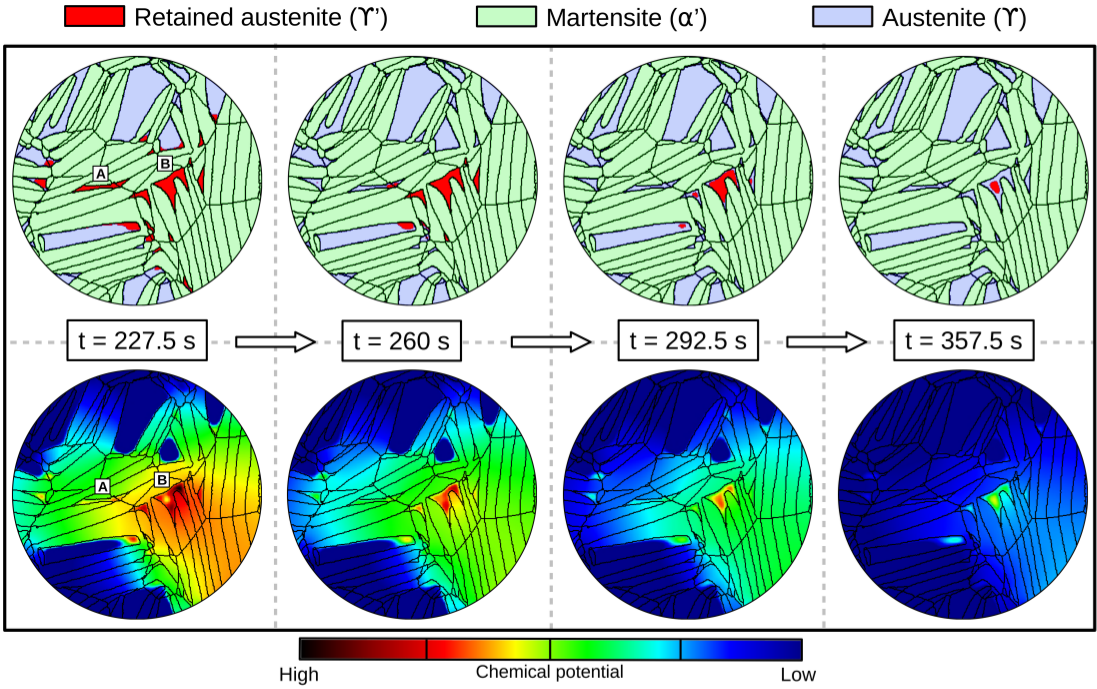}
    \end{tabular}
    \caption{ A section of the microstructure pertaining to Fig.~\ref{fig:009_60Evolution} which indicates the destabilisation of austenite \lq islands\rq \thinspace surrounded by martensite plates. The change in the chemical potential distribution which governs the partitioning through martensite is shown.
    \label{fig:island}}
\end{figure}
 
With increase in the volume fraction of martensite, the remnant phase gets enveloped between the $\alpha'$-plates forming austenite \lq islands\rq \thinspace~\cite{samanta2013development}.
A fraction of the microstructure, depicted in  Fig.~\ref{fig:009_60Evolution}, is sectioned to investigate the carbon enrichment in the austenite islands.
Fig.~\ref{fig:island} shows the distribution of carbon in and around the islands, $A$ and $B$, during the partitioning. 
The consideration of the chemical potential as the dynamic variable, in the phase-field model adopted to simulate the partitioning (Sec.~\ref{sec:part}), facilitates its graphical representation. 
The evolution of the potential distribution which dictates the carbon diffusion in the islands is illustrated in Fig.~\ref{fig:island}. 

Akin to other sections of the austenite, the concentration in the islands rises with the expulsion of carbon from the surrounding martensite. 
Owing to the size and the distribution, certain austenite islands like $A$ and $B$ become entirely stable at $t=227.5$~s, as illustrated in Fig.~\ref{fig:island}. 
As the partitioning proceeds, while the accumulation of the carbon is reduced by the diffusion within the austenite grains, the concentration decrease within the island is achieved by the migration of the carbon to the neighbouring austenite grains through the martensite. 
This reverse diffusion of the carbon, from the austenite island to the martensite and subsequently, to the destabilised austenite is initially observed in $A$ as shown in Fig.~\ref{fig:island} at $t=260$~s. 
A similar behaviour is later exhibited by the island $B$ at $t=357.5$~s. 
The chemical-potential distribution in Fig.~\ref{fig:island} indicates that the potential difference which determines the rate of the carbon diffusion is much less between the island and the surrounding martensite when compared to the potential difference within the 
austenite grain. 
In addition to the diffusion pathway through the martensite, this potential distribution protracts the destabilisation of the retained austenite in the island. 
In other words, the prolonged existence of the retained austenite in the $f^{\alpha'}=0.6$ microstructure, as presented in Fig.~\ref{fig:c009_v60}, is established by the combination of the increased carbon partitioning and the complex  low-rate destabilisation pathway adopted by the islands.

\begin{figure}
    \centering
      \begin{tabular}{@{}c@{}}
      \includegraphics[width=0.8\textwidth]{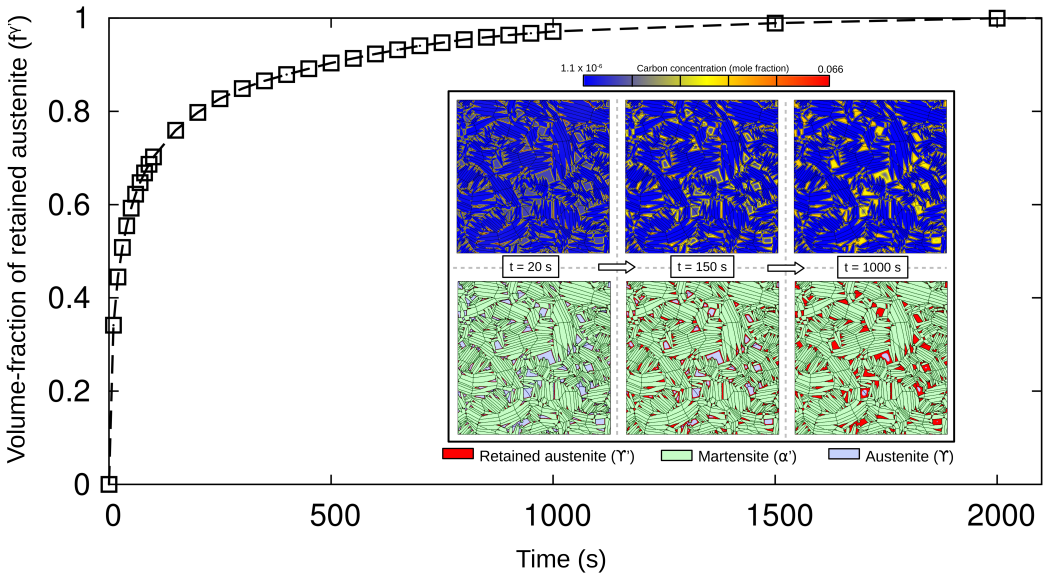}
    \end{tabular}
    \caption{ The temporal increase in the amount of retained austenite accompanying the carbon partitioning in the microstructure with very high martensite volume-fraction,  $f^{\alpha'}=0.8$. The change in the carbon distribution and the consequent stabilisation of the retained austenite in the polycrystalline structure is included as the subset of the plot.
    \label{fig:c009v80}}
\end{figure}

The temporal change in the volume fraction of retained austenite during partitioning in the microstructure with $80\%$ of martensite is illustrated in Fig.~\ref{fig:c009v80}. 
The carbon distribution which facilitates this increase in the retained austenite and the evolution of the $\gamma'$ in the multiphase system is included as a subset in Fig.~\ref{fig:c009v80}. 
With increase in the volume fraction of the martensite, the distribution of the remnant austenite gets increasingly dispersed, and enveloped by the martensite plates. 
As a result, the entire austenite volume in the microstructure evolves into separate islands like $A$ and $B$ as illustrated in Fig.~\ref{fig:island}.
This configuration of the austenite simplifies the migration of carbon during partitioning. 
Furthermore, although the $c^{\alpha '}_\text{CCE}$ increases with the martensite volume-fraction, the amount of carbon diffusing to the austenite is significantly higher in this system owing to the increased volume of the $\alpha'$. 
While the intense carbon enrichment in the austenite favours the stabilisation, its configuration in the polycrystalline set-up prolongs any diffusion, which facilitates the destabilisation. 
As discussed above, these aspects of the carbon partitioning collectively contribute to the observed increase in the retained austenite volume-fraction and complete transformation of $\gamma \rightarrow \gamma'$. 
In addition to the absolute stabilisation of the austenite, the trend exhibited in  Fig.~\ref{fig:c009v80} signifies a sudden increase in $f^{\gamma'}$ governed by the localized accumulation of the carbon and the subsequent diffusion resulting in the eventual expansion of the retained austenite. 

\subsubsection{High carbon concentration: $\textnormal{Fe-0.5}~\textnormal{wt}\%\textnormal{C}$ ($c^{\textnormal{alloy}}=0.0225$)}

\begin{figure}
    \centering
      \begin{tabular}{@{}c@{}}
      \includegraphics[width=0.75\textwidth]{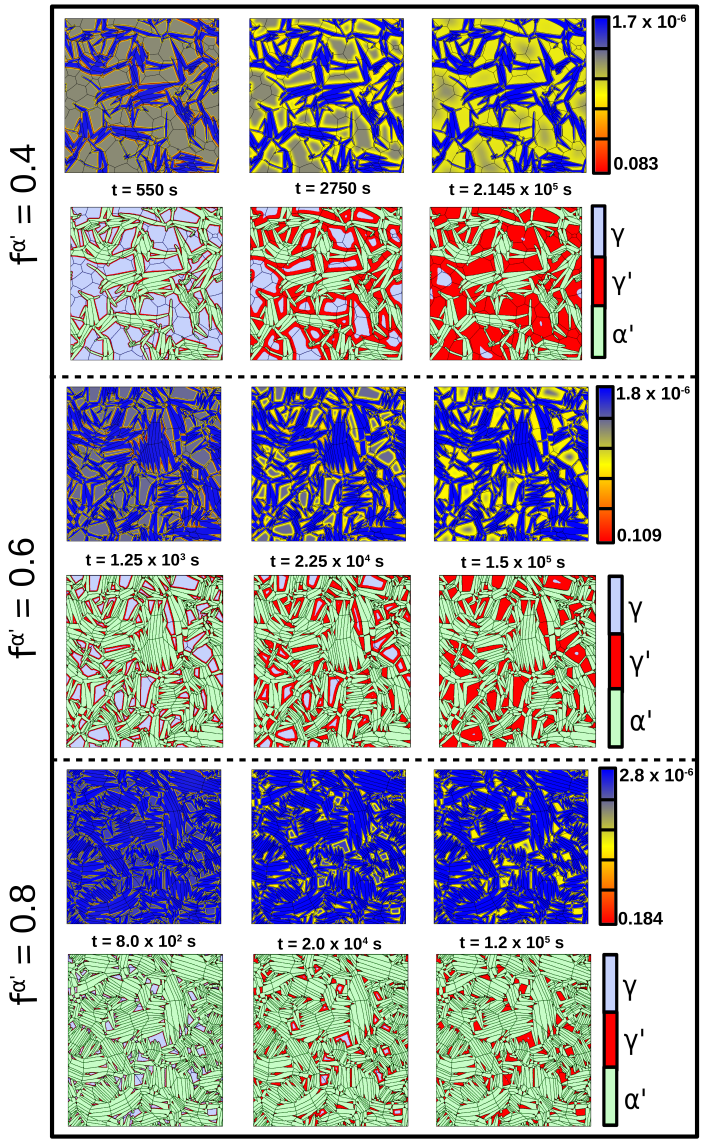}
    \end{tabular}
    \caption{ The carbon partitioning in an Fe-C system with $0.5~\text{wt}\%$ of carbon in the quenched microstructure with different martensite volume-fraction, $f^{\alpha'}=0.4$, $0.6$ and $0.8$. The evolution of the carbon in relation to the phase distribution present the consequent stabilisation of the austenite in the polycrystalline set-up. 
    \label{fig:c02_evolve}}
\end{figure}

The quenched microstructure, employed in the Sec.~\ref{sec:low_c}, is adopted to investigate the stability of the austenite during the partitioning in the binary system of high carbon concentration, Fe-0.5~\text{wt}\%C.
In Fig.~\ref{fig:c02_evolve}, the diffusion of carbon accompanying the partitioning at $700~\degree \text{C}$ and the resulting change in the distribution of the retained austenite is collectively presented  for all the three microstructures with $f^{\alpha'}=0.4$, $0.6$ and $0.8$. 
The change in the CCE-composition, $c^{\gamma}_\text{CCE}$ and $c^{\alpha '}_\text{CCE}$, with increase in the initial composition of the system is tabulated in Table~\ref{table:high_c}.
\begin{table}[!ht]
\centering
\caption{CCE composition of martensite and austenite in Fe-C system of $0.0225$ mole frac. carbon.} 
\scalebox{0.9}{
\begin{tabular}{l|l|l} 
\hline
{$f^{\alpha'}$} & { $c^{\gamma}_\text{CCE}$ (mole frac.)} & { $c^{\alpha '}_\text{CCE}$ (mole frac.)}  \\ \hline	
0.4	& $1.7 \times 10^{-6}$  	&  0.035\\ \hline
0.6	& $1.8 \times 10^{-6}$  	&  0.052\\ \hline
0.8	& $2.8 \times 10^{-6}$  	& 0.0882\\ \hline
\end{tabular}
}
\label{table:high_c}
\end{table}
When compared to Sec.~\ref{sec:low_c}, the increase in the composition $c^{\text{alloy}}$ favours the complete stabilisation of the austenite in $f^{\alpha'}=0.4$ distribution, despite the low martensite volume-fraction.
Furthermore, the entire austenite volume stabilises in all the polycrystalline structures, irrespective of the martensite volume-fraction ($f^{\alpha'}$). 
In Fig.~\ref{fig:c02_evolve}, however, a significant difference in the rate of the stabilisation is evident.
Noticeably, the austenite in the $f^{\alpha'}=0.8$ microstructure becomes stable at a faster rate when compared to $f^{\alpha'}=0.4$.
\begin{figure}
    \centering
      \begin{tabular}{@{}c@{}}
      \includegraphics[width=1.0\textwidth]{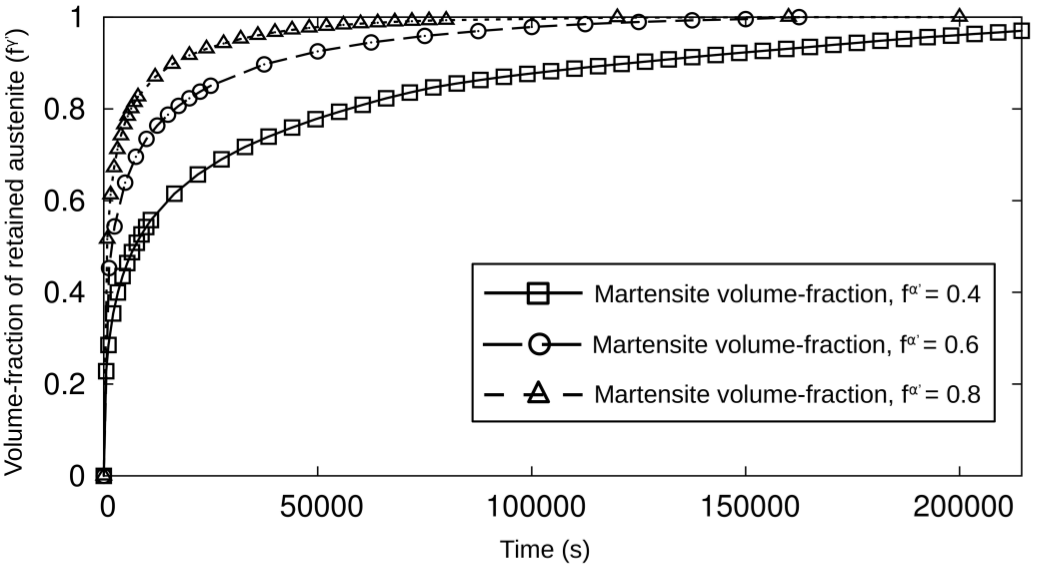}
    \end{tabular}
    \caption{ The change in the retained austenite volume-fraction ($f^{\gamma'}$) with the partitioning in the microstructures with varying martensite volume, $f^{\alpha'}=0.4$, $0.6$ and $0.8$.  
    \label{fig:c02}}
\end{figure}
To illustrate the kinetics of the austenite stabilisation, the temporal change in $f^{\gamma'}$ is plotted in Fig.~\ref{fig:c02}.
The trend in the evolution of the $f^{\gamma'}$ is independent of the martensite volume-fraction and is identical for all microstructures. 
Moreover, this scheme in the temporal change of $f^{\gamma'}$ is similar to Fig.~\ref{fig:c009v80}. 
Due to the substantial difference in the diffusivities and the CCE endpoints, the formation of the stable austenite begins with the accumulation of the carbon in the $\alpha'\gamma$-interface, irrespective of the alloy composition and the martensite volume-fraction. 
In all microstructures, this carbon accumulation accounts for the drastic increase in $f^{\gamma'}$ in the initial stages of the partitioning. 
Visible deviations between the different microstructures in the evolution of the retained austenite is introduced in the subsequent stages of the partitioning, as shown in Fig.~\ref{fig:c02_evolve}.
The variation in the kinetics of the stabilisation is governed by the carbon diffusion within the austenite grains and is therefore, dictated by the austenite volume-fraction.
Correspondingly, the austenite in $f^{\alpha'}=0.8$ microstructure becomes stable in fewer time-steps when compared to the other polycrystalline system, $f^{\alpha'}=0.4$ and $0.6$. 
Since,  the complex diffusion adopted by the islands (Fig.~\ref{fig:island}) is not expected with the increase in alloy concentration, the difference in the stabilisation kinetics is  influenced entirely by the martensite volume-fraction.

\section{Conclusion}

The mechanical properties of steel, like toughness and ductility, are enhanced by the presence of austenite~\cite{bhadeshia1979tempered,ojima1998application}.
Austenite, however, becomes increasingly unstable at low temperatures and more importantly, decomposes at room temperature.
Alloying elements, referred to as austenite stabilisers, are often added to sustain the austenite at low temperatures.
Of the different additives, carbon is considered as one of the potent alloying elements to stabilise austenite.
A recently developed heat-treatment technique, called Quenching and Partitioning, employs carbon as the stabiliser to retain austenite at room temperature~\cite{speer2003carbon}.
The Q\&P processing involves quenching of fully or partially austenized steels to yield a microstructure with the combination of austenite and martensite in desired volume fractions.
To facilitate the diffusion of carbon from martensite to austenite, the temperature is increased during the partitioning.
In some instances, however, the partitioning is achieved by holding the alloy at the quenching temperature.
Since the distribution of the alloys varies significantly from the equilibrium volume-fraction of the phases, and no interface migration is assumed, a deviation from the equilibrium composition of the phases is predicted, and an endpoint for the partitioning is ascertained from the carbon activity.
The corresponding compositions are referred to as constrained carbon equilibrium (CCE) concentrations, $c_{\text{CCE}}^{\Theta}$, where $\Theta$ can be austenite ($\gamma$) or martensite ($\alpha'$).

In the present work, the phase transformation accompanying the quenching and the carbon diffusion during the partitioning is simulated by adopting the phase-field approach.
The diffusionless martensite transformation in a polycrystalline microstructure is simulated using an elastic model, which has already been shown to recover the sharp-interface solutions~\cite{schneider2017small,schoof2017multiphase}.
Three different volume fractions of martensite are arbitrarily chosen and the corresponding quenching temperature is determined analytically.
The chemical driving force from the CALPHAD database~\cite{gustafson1985thermodynamic} is incorporated and the predicted volume-fraction of the martensite is achieved.

Before simulating the carbon partitioning, the free-energy curve from the CALPHAD database is modified to reproduce the constrained carbon equilibrium condition.
This modified data is incorporated in an established phase-field model, which is known for its quantitatively modelling of solidification~\cite{hotzer2015large}, solid-state transformation~\cite{mushongera2018phase} and shape-instabilities~\cite{amos2018globularization}, to simulate the carbon diffusion during partitioning.
Based on the distribution of carbon, the potential retained-austenite is determined from the analytical relation and its evolution is investigated.
Furthermore, to capture the kinetics of the partitioning and thereby, the evolution of the retained austenite, diffusivities governed by the temperature and composition are employed. 
In addition to the different martensite volume-fraction, two alloy compositions of varying carbon concentrations are considered.

The considerable difference in the diffusivities and the partitioning endpoints, $c_{\text{CCE}}^{\gamma}$ and $c_{\text{CCE}}^{\alpha'}$, lead to the accumulation of carbon in the austenite grains along the $\alpha'\gamma$-interface. 
This carbon accumulation in the early stages of the partitioning enhances the stabilisation of the austenite adjacent to the $\alpha'\gamma$-interface, independent of the composition or the martensite volume-fraction considered in the current study. 
However, as the partitioning proceeds, the accumulation diminishes through the carbon diffusion within the austenite which destabilises the austenite in certain microstructures.
This austenite destabilisation in the specific microstructure indicates the pivotal influence of the martensite volume-fraction and the alloy composition. 
Furthermore, it is identified that the distribution of the phases influences the kinetics of the austenite stabilisation (or destabilisation).
Since the diffusion from the carbon-enriched austenite island proceeds through the martensite, the destabilisation is extensively prolonged. 
In the current work, it is shown that the entire austenite volume gets stabilised with an increase in the carbon concentration, although visible differences are observed in the kinetics.
Despite the difference in the time taken for the stabilisation, owing the carbon accumulation and the subsequent diffusion within austenite, the trend in the temporal evolution of the retained austenite remains unaltered. 

Since the primary focus of this work is to render a  polycrystalline investigation of the quenching and partitioning which adopts the CCE conditions, the interface migration during the partitioning under the influence of the elasticity  is overlooked.
However, attempts are made to report on the relaxation of the strains, and corresponding interface migration, during the carbon diffusion from martensite. 
Furthermore, no significant difference is made between the volume and interface diffusion. 
Since a considerable change in the kinetics of the retained-austenite evolution is expected from the dominance of the interface diffusion, this aspect of the partitioning will be addressed in the near future. 
This study considers the fully-austenized polycrystalline structure as the starting point for the Q\&P processing, whereas partial austenization is also adopted industrially. 
Therefore, the martensite transformation and carbon partitioning in a partially austenized microstructure with the ferrite grains, under CCE conditions, will be examined in the upcoming works.
Although the influence of the alloying elements, which prevent the formation of carbides and other ill-suited phases, during the martensite transformation and carbon partitioning has not been analysed in this work, attempts are made to address this aspect of the Q\&P technique in the subsequent studies.

\section{Appendix}

The lattice parameters of the phases considered in the present work, although not explicitly, play a critical role. 
The Koistinen-Marburger relation predicts the volume fraction of the martensite for given quenching temperature. 
However, the calculation of the CCE-concentrations, $c^{\gamma}_\text{CCE}$ and $c^{\alpha '}_\text{CCE}$, assumes mole fractions of the constituent phases.
Often, the molar volume facilitating the conversion of volume fractions to mole fractions is assumed to be constant for the martensite and austenite. 
In the present work, an accurate calculation of the molar volume is made from the lattice parameters, which are sensitive to the carbon concentration~\cite{rammo2006model}. 
The lattice parameter $a_{\gamma}$ of the austenite is determined from 
\begin{align}\label{eq:a_austenite}
 a_{\gamma}= 2\sqrt{2}r_{\text{Fe}}+\Bigg[\frac{2(r_{\text{Fe}}+r_{\text{C}})-2\sqrt{2}r_{\text{Fe}}}{4 \times 10}\Bigg]X_{\text{C}},
\end{align}
where $r_{\text{Fe}}$ and $r_{\text{C}}$ are the respective atomic radii of iron and carbon atoms, and $X_{\text{C}}$ is the number of carbon atoms for $100$ iron atoms.
The two lattice parameters of the body-centered tetragonal martensite, $a_{\alpha'}$ and $c_{\alpha'}$, are calculated using the relation
\begin{align}\label{eq:a_martensite}
 a_{\alpha'}= \frac{4r_{\text{Fe}}}{\sqrt{3}}-\Bigg[\frac{\frac{4r_{\text{Fe}}}{\sqrt{3}}-\sqrt{2}(r_{\text{Fe}}+r_{\text{C}})}{4 \times 10}\Bigg]X_{\text{C}}
\end{align}
and 
\begin{align}\label{eq:c_martensite}
 c_{\alpha'}= \frac{4r_{\text{Fe}}}{\sqrt{3}}+\Bigg[\frac{2(r_{\text{Fe}}+r_{\text{C}})-\frac{4r_{\text{Fe}}}{\sqrt{3}}}{4 \times 10}\Bigg]X_{\text{C}}.
\end{align}
The molar volume, sensitive to the concentration, is determined from the unit cells, by relating the volume occupied by one unit-cell and the number of unit cells in one mole of the phase.

\section*{Acknowledgments}

PGK Amos thanks the financial support of the German Research Foundation (DFG) under the project AN 1245/1. 
Ephraim Schoof acknowledges the financial support of Ministry of Science, Research and Arts of Baden-Wuerttemberg, under the grant 33-7533.-30-10/25/54.
The authors are also grateful for the editorial support by Leon Geisen.
This work was performed on the computational resource ForHLR II, funded by the Ministry of Science, Research and Arts of Baden-Wuerttemberg and the DFG.

\section*{References}

\bibliographystyle{elsarticle-num}
\bibliography{library.bib}
\end{document}